\documentclass[fleqn,10pt]{wlscirep}
\usepackage[utf8]{inputenc}
\usepackage[T1]{fontenc}
\usepackage{physics}
\usepackage{amsmath}
\usepackage{subfiles}
\usepackage{subcaption}
\usepackage{float}
\usepackage{bm}
\DeclareMathOperator{\spn}{\mathrm{span}}

\newcommand{\myUnder}[2]{{#1}_{#2}}
\newcommand{\myTab}[5][H]{
\begin{table}[#1]
    \caption{#2}
    \label{#3}
    \begin{center}
    \begin{tabular}{#4}\hline
    #5
    \end{tabular}
    \end{center}
\end{table}%
}

\title{Asymmetric quantum decision-making}

\author[1,*]{Honoka Shiratori}
\author[1]{Hiroaki Shinkawa}
\author[1]{André Röhm}
\author[1]{Nicolas Chauvet}
\author[3]{Etsuo Segawa}
\author[2]{Jonathan Laurent}
\author[2]{Guillaume Bachelier}
\author[1]{Tomoki Yamagami}
\author[1]{Ryoichi Horisaki}
\author[1]{Makoto Naruse}
\affil[1]{The University of Tokyo, Department of Information Physics and Computing, Graduate School of Information Science and Technology, Tokyo, 113-8656, Japan}
\affil[2]{Université Grenoble Alpes, CNRS, Institut Néel, Grenoble, 38042, France}
\affil[3]{Yokohama National University, Graduate School of Environment and Information Sciences, Yokohama, 240-8501, Japan}

\affil[*]{8308123768hs@g.ecc.u-tokyo.ac.jp}

\begin{abstract}
Collective decision-making is crucial to information and communication systems.
Decision conflicts among agents hinder the maximization of potential utilities of the entire system.
Quantum processes can realize conflict-free joint decisions among two agents using the entanglement of photons or quantum interference of orbital angular momentum (OAM).
However, previous studies have always presented symmetric resultant joint decisions.
Although this property helps maintain and preserve equality, it cannot resolve disparities.
Global challenges, such as ethics and equity, are recognized in the field of responsible artificial intelligence as responsible research and innovation paradigm.
Thus, decision-making systems must not only preserve existing equality but also tackle disparities.
This study theoretically and numerically investigates asymmetric collective decision-making using quantum interference of photons carrying OAM or entangled photons.
Although asymmetry is successfully realized, a photon loss is inevitable in the proposed models.
The available range of asymmetry and method for obtaining the desired degree of asymmetry are analytically formulated.
\end{abstract}

\begin{document}

\flushbottom
\maketitle
%
%
\thispagestyle{empty}


\section*{Introduction}

Even in situations with limited knowledge, people are required to make decisions by estimating and believing which choice is profitable \cite{daw2006cortical}.
The multi-armed bandit problem model depicts the decision-making process in uncertain environments, wherein each player is assumed to intend to maximize reward by predicting the best one among several slot machines, referred to as arms, whose reward probabilities are unknown \cite{sutton2018reinforcement}.
In a multi-armed bandit problem, exploration is necessary to predict reward probabilities precisely; however, excessive exploration can diminish the sum of obtained rewards \cite{Auer,march1991exploration}, whereas minimal explorations can result in the best arm being missed.
Furthermore, when numerous players are engaged in the game, the problem is referred to as a competitive multi-armed bandit problem \cite{chauvet}.
In this case, decision conflicts are another problem because multiple players choosing the same arm can result in a bottleneck and consequently impede the profits of the entire group \cite{Lai, Kim}.

Quantum approaches have been extensively studied to solve uncertain problems \cite{
steinbrecher2019quantum,
saggio2021experimental,
flamini2020photonic,
bukov2018reinforcement,
niu2019universal,
porotti2019coherent}. 
The quantum properties of photons can aid in solving the problem of decision conflicts in collective decision-making \cite{chauvet,amakasu,shinkawa2022conflict}.
Two previous studies developed quantum systems enabling conflict-free decision-making between two players.
The first study utilized the Hong-Ou-Mandel effect of orbital angular momentum (OAM) \cite{amakasu,shinkawa2022conflict}, whereas the second one utilized entangled photons \cite{chauvet}.

\begin{figure}[t]
\centering
\includegraphics[width=\textwidth]{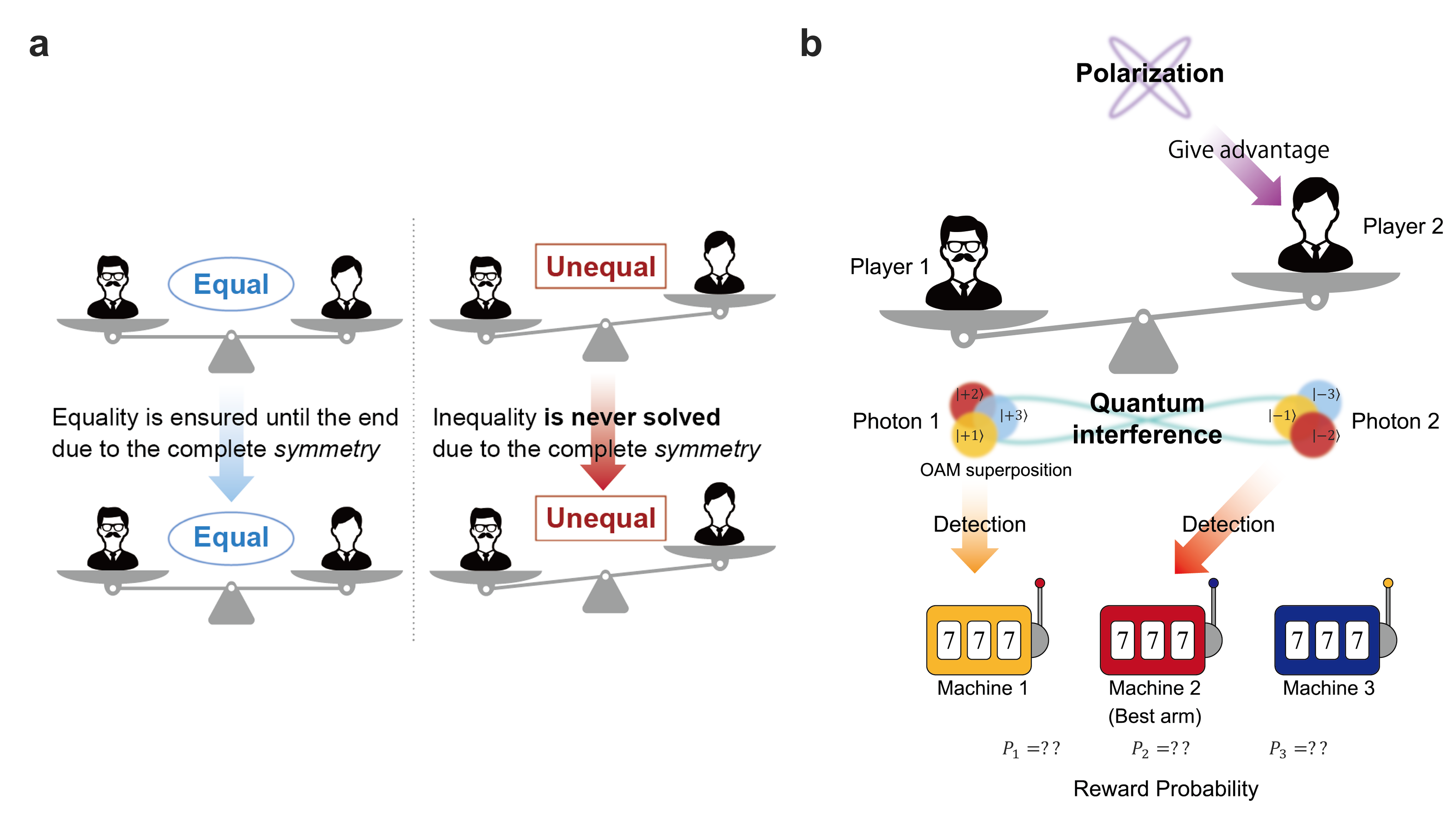}
\caption{(a) The necessity of asymmetric treatments. The symmetric OAM system in the previous study can maintain the existing equality but cannot solve inequality. (b) Stochastic detection of OAM corresponds to a probabilistic selection of a player. Polarization enables us to diminish inequalities between players.}
\label{concept}
\end{figure}

However, these systems prohibit conducting affirmative actions  
\cite{holzer2000assessing} to reduce disparities between players, primarily  because the decisions made are always symmetric.
Namely, the probability of player X selecting arm $l$ and player Y choosing arm $m$ is inevitably the same as that of player X selecting arm $m$ and player Y choosing arm $l$.
This property is referred to as \textit{symmetry},
owing to which both players are always treated evenly; essentially, equality is ensured \cite{bolton2000erc}. 
We refer to the previous study utilizing the Hong-Ou-Mandel effect of OAM as the symmetric OAM system, and the other one utilizing entangled photons as the symmetric entangled photon decision maker.
The symmetric property is suitable when players are equal since the beginning of the game because, on average, equality is ensured at all times by symmetry.
However, consider if one player is in a much more advantageous position compared with the other prior to the game; this inequality cannot be resolved by the aforementioned systems owing to symmetry (Figure \ref{concept}a).
Thus, these previous systems are superior in maintaining equality; however, they cannot reduce disparities.

To facilitate affirmative actions in resolving inequalities, decision-making must be \textit{asymmetric} such that a disadvantaged person or entity is more likely to choose the better arm than an advantaged person or entity.
Asymmetry is the property that allows the probability of player X selecting arm $l$ and player Y choosing arm $m$ to differ from that of the opposite case.
Previously established systems enabled only symmetric treatments, whereas the decision-making systems proposed in this study can control asymmetry by enabling asymmetric treatments, thus being able to facilitate advantageous outcomes for underprivileged agents (Figure \ref{concept}b).
Note that the initial aforementioned disparities are recognized in various serious social issues ranging from earning differentials, gender gaps, and educational inequalities \cite{blau1994rising,shen2013inequality,sandel2020tyranny,breen2005inequality}.
In addition, the importance of focusing on the wider context of global challenges, such as ethics and fairness, is recognized in the field of responsible artificial intelligence (AI) as responsible research and innovation paradigm (RRI)\cite{arrieta2020explainable, Stahl2018}.
Thus, considering the social context and RRI, 
ensuring existing equality may be insufficient, and affirmative actions must be enabled to diminish disparities.
Another context is setting priority in information and communication services. 
Prioritized agents or entities should receive higher rewards than others while avoiding decision conflicts. 

This study proposed improvements in quantum models and incorporated the potential to address disparities by realizing asymmetric decision-making and enabling control of asymmetry in the competitive multi-armed bandit problem.
First, a quantum model was proposed by applying the Hong-Ou-Mandel effect with polarization dependencies, which is referred to as the \textit{asymmetric} OAM system. 
This corresponded to an enhanced version of the symmetric OAM system proposed by Amakasu et al. \cite{amakasu} by further incorporating the polarization-dependent effects. 
Next, the achievable asymmetric decision-making range was clarified analytically. 
Furthermore, two models to be compared with the asymmetric OAM system were investigated.
One was the extension of the symmetric entangled photon decision-maker\cite{chauvet}, whereas the other was an extension of the symmetric OAM system \cite{amakasu}. 
The proposed asymmetric OAM system can provide asymmetric decision-making with negligible photon loss, provided the intended asymmetry is significant, whereas the entangled-photon approach suffers from significant photon loss. 
Conversely, the proposed asymmetric OAM system must accompany photon loss or decision conflicts when the decision is required to be symmetric, whereas the entangled photon approach accomplishes negligible photon loss in the corresponding situation. 
Thus, a trade-off exists between the proposed asymmetric OAM system and the entangled photon system.

\section*{Asymmetric decision-making by OAM}
\begin{figure}[tb] 
\centering
\includegraphics[width=\textwidth]{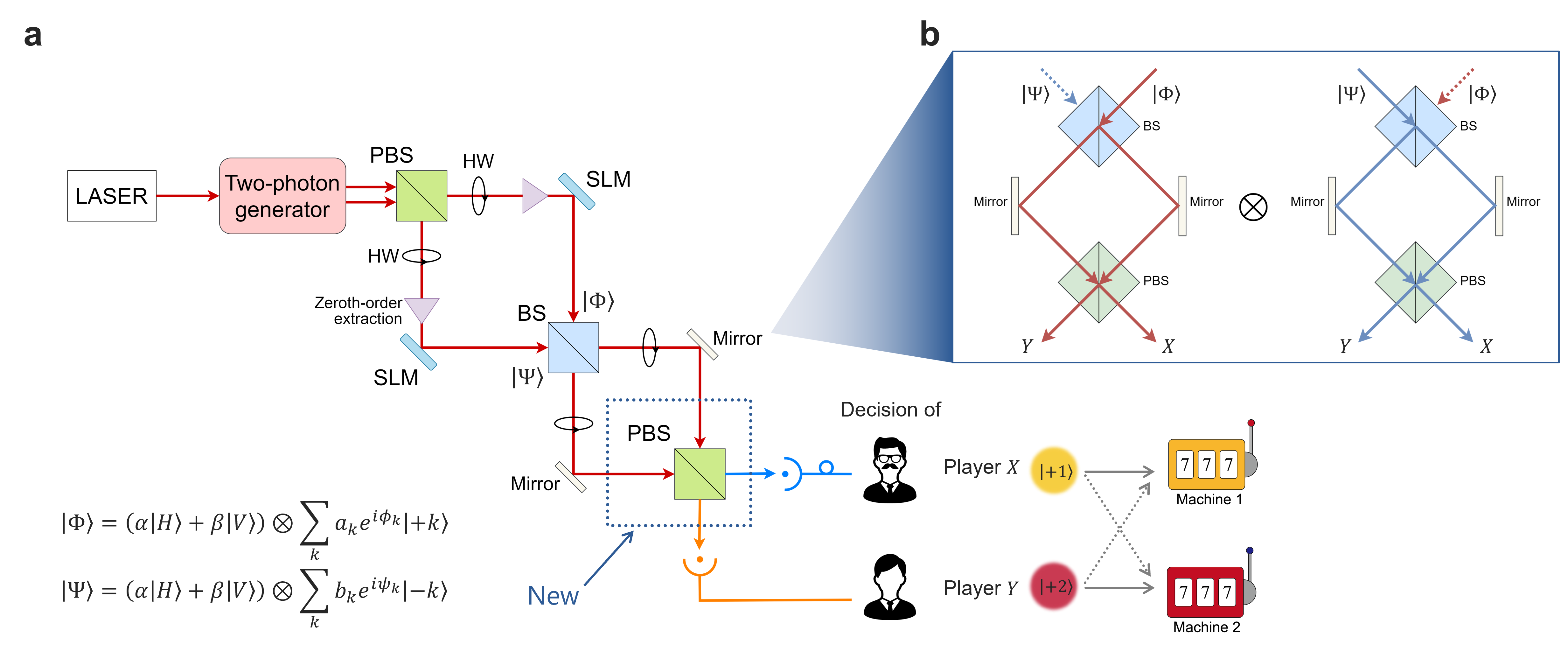}
\caption{(a) System architecture for the asymmetric OAM decision making. 
It differs from the symmetric OAM system in that PBS is added, and the basis of polarization is added to the photon state. (b) Schematic illustration of the paths of photons.}
\label{OAM_setting2}
\end{figure}

This section proposes the manner in which asymmetric decision-making can be realized by the decision-making system utilizing OAM.
Figure \ref{OAM_setting2}a shows the construction of the system for the two-player-$K$-armed bandit problem.
The OAM detected at X corresponds to the arm selected by player X, whereas that detected at Y corresponds to the arm selected by player Y.
Two inputs $\Phi$ and $\Psi$ are represented by two bases: OAM and polarization.
This system is different from that in the previous study \cite{amakasu} in that one polarization beam splitter (PBS) is added to it, and photons have polarizations. The polarization of photons is represented by $\alpha$ and $\beta$, i.e., the probability amplitudes of photons having horizontal and vertical polarizations are $\alpha$ and $\beta$, respectively, and the relation $|\alpha|^2 + |\beta|^2 = 1$ holds.
Note that $\alpha,~\beta \in \mathbb{R}$ in this thesis.
$\alpha$ and $\beta$ are represented by $\cos\theta$ and $\sin\theta$ respectively later in this thesis.

\subsection*{Formulation}

This section provides a mathematical derivation of the probabilities corresponding to pairs of decisions based on the system presented in Figure \ref{OAM_setting2}a for the case involving $K$ choices, that is, $K$ OAMs.

First, we give the space to treat polarization and OAMs. Let us denote the horizontal and vertical polarization states by
\begin{equation}
    \ket{H}=\left[\begin{array}{c}
     1  \\
     0 
    \end{array}\right], ~~
    \ket{V}=\left[\begin{array}{c}
     0  \\
     1 
    \end{array}\right],
\end{equation}
respectively. Then the Hilbert space corresponding to polarization states is described as
\begin{equation}
    \mathcal{H}_p := \spn\{\ket{H},\ \ket{V}\} \simeq \mathbb{C}^2.
\end{equation}

OAM states are represented by integers; their signs and absolute values denote directions (right $(+)$- or left $(-)$-handed) and numbers of  intertwined
helices, respectively \cite{Yao}.
Especially, the numbers of intertwined
helices are utilized to identify the selected arms; for example, OAM $\pm k$ corresponds to arm $k$.
Thus, we limit the possible values of OAMs to $\pm 1,\ \pm 2,\ \cdots \pm K$.
Here, for $k\in [K] := \{1,\,2,\,\cdots,\,K\}$, we define vector $\ket{\pm k}\in \mathbb{C}^{2K}$ corresponding to OAM $\pm k$ as follows:
\begin{equation}
    \ket{+k} = [~0,\dots,\stackrel{\text{$k$}}{\breve{1}},\dots,0~]^\top ,~~
    \ket{-k} = [~0,\dots,\stackrel{\text{$k+K$}}{\breve{1}},\dots,0~]^\top,
\end{equation}
where superscript $\top$ on a matrix represents the transpose of the matrix. Then the Hilbert space corresponding to OAM states is described as
\begin{equation}
    \mathcal{H}_o:= \spn \{\ket{\ell}\,|\,\ell\in \pm [K]\} \simeq \mathbb{C}^{2K},
\end{equation}
where $\pm[K] = \{\pm 1,\,\pm 2,\,\cdots,\,\pm K\}$. As polarization states and OAM states are independent, the hybrid states are in the composite Hilbert space defined as
\begin{align}
    \mathcal{H}_s = \mathcal{H}_p \otimes \mathcal{H}_o = \spn\{\ket{P}\otimes \ket{\ell}\,|\,P\in\{H,\ V\},\ \ell\in \pm[K]\} \simeq  \mathbb{C}^{4K},
\end{align}
as in [\citeonline{Vallone2014}].

The hybrid states of OAM and polarization can be generated using spatial light modulators (SLMs). The first input is represented as:
\begin{equation}
\label{eq:phi1}
    \Phi = \left[ \begin{array}{c}
         \alpha \\
         \beta
    \end{array} \right] \otimes ~~ \Tilde{\Phi},~~ 
    \Tilde{\Phi}:= \sum_{k=1}^{K} a_k e^{i\phi_k}\ket{+k} =  
    [  ~a_1 e^{i\phi_1}, ~\cdots, ~a_k e^{i\phi_k}, ~\cdots, ~a_K e^{i\phi_K} ,\overbrace{~0, ~\cdots, ~0}^{\text{$k$ elements}} ~]^\top \in \mathcal{H}_o,
\end{equation}
where $a_k \in \mathbb{R}$ and $\phi_k\in [0,\ 2\pi)$ for all $k\in [K]$, and the equation $\Sigma_{k=1}^K a_k^2 =1$ holds.
The elements $\alpha$ and $\beta$, which represent a vector in $\mathcal{H}_p$, are the probability amplitudes of photon $\Phi$ with horizontal and vertical polarizations, respectively.
The elements in the latter half of $\Tilde{\Phi}$ were all zero because $\Phi$ was designed to have only positive OAM. 
In the previous research, one player manipulated $\Tilde{\Phi}$ according to his or her preference.
$\Phi$ is a $4K$ dimensional vector because of the tensor product.
The first $2K$ elements of $\Phi$ correspond to the probability amplitudes of horizontal polarization, consisting of the OAMs.
Essentially, the squared sum of the first $2K$ elements of $\Phi$, $ \left[~ \alpha, ~~0 ~\right]^\top \otimes \Tilde{\Phi} $, is $|\alpha|^2$, which is the probability that a photon $\Phi$ exhibits horizontal polarization.
In addition, the latter $2K$ elements, $\left[~ 0, ~~\beta ~\right]^\top \otimes \Tilde{\Phi} $, are the probability amplitudes of vertical polarization.

After passing the first beam splitter, $\Phi$ is transformed to:
\begin{equation}
    \Phi'= \left(I_2 \otimes  A\right)  \Phi ,~~ A :=  \sum_{\ell\in\pm[K]}\frac{1}{\sqrt{2}}\Bigl(\ketbra{\ell}{\ell} + i\ketbra{-\ell}{\ell}\Bigr) = 
    \frac{1}{\sqrt{2}} \left[ \begin{array}{cc}
         I_K & iI_K  \\
         iI_K & I_K
    \end{array} \right] ,
\end{equation}
where $A$ corresponds to the effect of a beamsplitter in the $2K$-dimensional Hilbert space of OAM, $\mathcal{H}_o$.
Here, for $N \in \mathbb{N}$,  $I_N$ indicates a $N$ by $N$ identity matrix.  
Essentially, OAM did not change if a photon transmits through a beam splitter, whereas the sign of OAM reversed, and the probability amplitude was multiplied by $i$ if it was reflected. 
However, because both the OAM and polarization were considered herein, the effect of a beam splitter on photon states was $I_2 \otimes A$, which performs a unitary transformation on $\mathcal{H}_s$.
Subsequently, based on the reflection at mirrors after the beam splitter, $\Phi'$ is transformed to:
\begin{equation}
    \Phi'' = \left( I_2 \otimes R \right) \Phi' = \left(I_2 \otimes RA \right) \Phi ,~~
    R := \sum_{\ell\in\pm[K]} i\ketbra{-\ell}{\ell} = 
    \left[ \begin{array}{cc}
         O_K & iI_K \\
         iI_K & O_K
    \end{array} \right],
\end{equation}
wherein $R$ corresponds to the effect of the reflection by a mirror in the $2K$-dimensional Hilbert space of OAM, and
$O_N$ implies a $N$ by $N$ zero matrix. 
$R$ implies that the reflection reverses the signs of OAMs, and probability amplitudes are multiplied by $i$.
However, because the polarization must be considered in addition to OAM, the effect of reflections on photon states should be $I_2 \otimes R$, which performs a unitary transformation on $\mathcal{H}_s$.

At a polarization beam splitter, OAM with horizontal polarization is transmitted, whereas OAM with vertical polarization is reflected and multiplied by $i$.
Therefore, the effect of a polarization beam splitter, which acts on $\mathcal{H}_s=\mathcal{H}_p \otimes \mathcal{H}_o$, can be represented by the following $4K$ by $4K$ matrix $C$:
\begin{equation}
    C := \ketbra{H}{H} \otimes I_{2K} + \ketbra{V}{V}\otimes \sum_{\ell\in\pm[K]} i\ketbra{-\ell}{\ell}
    = \left[\begin{array}{cc}
       I_{2K}  &  O_{2K}  \\
       O_{2K}  &  i\sigma \otimes I_K
    \end{array} \right], ~~ \sigma := \left[ \begin{array}{cc}
       0  &  1\\
       1  &  0
    \end{array} \right].
\end{equation}
The operator $C$ performs a unitary transformation on $\mathcal{H}_s$.
Because detectors are sensitive only to OAMs, herein, we should consider the corresponding map $\mathcal{H}_s \rightarrow \mathcal{H}_o$ which is represented by a $2K$ by $4K$ matrix $\left[ \begin{array}{ccc}  I_{2K} & \Big| & I_{2K} \end{array} \right]$.
The probability amplitude of OAM $k$ and horizontal polarization and that of OAM $k$ and vertical polarization were added.
Therefore, the effect of the asymmetric OAM system on one input is expressed as:
\begin{equation}
    V := \left[ \begin{array}{ccc}  I_{2K}  &  \Big| &  I_{2K}    \end{array} \right] C \left( I_2 \otimes RA \right) 
    = \frac{1}{\sqrt{2}} \left[ \begin{array}{cccc}
       -I_K  & iI_K & -I_K  & -iI_K  \\
        iI_K & -I_K & -iI_K & -I_K
    \end{array} \right].
\end{equation}
Hence, the observed output of input $\Phi$ is:
\begin{equation}
    \Phi_{\mathrm{out}} 
    = V \Phi = V  \left[ \begin{array}{c}
        \alpha a_1 e^{i\phi_1} \\
        \vdots \\
        \alpha a_K e^{i\phi_K} \\
        0_K \\
        \beta a_1 e^{i\phi_1} \\
        \vdots \\
        \beta a_K e^{i\phi_K} \\
        0_K
    \end{array} \right] = \frac{1}{\sqrt{2}} \left[\begin{array}{c}
         -(\alpha + \beta) a_1 e^{i\phi_1}  \\
         \vdots \\
         -(\alpha + \beta) a_K e^{i\phi_K}  \\
         i(\alpha - \beta) a_1 e^{i\phi_1}  \\
         \vdots \\
         i(\alpha - \beta) a_K e^{i\phi_K} 
    \end{array} \right].
    \label{phiout}
\end{equation}
Note that $0_N$ indicates a zero vector with $N$ elements. 

Similarly, by considering the second input having the same polarization and OAMs with only negative signs:
\begin{equation}
\label{eq:psi1}
    \Psi = \left[ \begin{array}{c}
         \alpha \\
         \beta
    \end{array} \right] \otimes ~~ \Tilde{\Psi},~~
    \Tilde{\Psi}:= \sum_{k=1}^{K} b_k e^{i\psi_k}\ket{-k} = 
    [~\overbrace{0,~\cdots, ~0}^{\text{$k$ elements}},~ b_1 e^{i\psi_1}, ~\cdots, ~ b_k e^{i\psi_k}, ~ \cdots, ~ b_K e^{i\psi_K} ~]^\top,
\end{equation}
\begin{equation}
    \Psi_{\mathrm{out}} = V \Psi = V  \left[ \begin{array}{c}
        0_K \\
        \alpha b_1 e^{i\psi_1} \\
        \vdots \\
        \alpha b_K e^{i\psi_K} \\
        0_K \\
        \beta b_1 e^{i\psi_1} \\
        \vdots \\
        \beta b_K e^{i\psi_K}
    \end{array} \right] = \frac{1}{\sqrt{2}} \left[\begin{array}{c}
         i(\alpha - \beta) b_1 e^{i\psi_1}   \\
         \vdots \\
         i(\alpha - \beta) b_K e^{i\psi_K}   \\
         -(\alpha + \beta) b_1 e^{i\psi_1} \\
         \vdots \\
         -(\alpha + \beta) b_K e^{i\psi_K} 
    \end{array} \right].
    \label{psiout}
\end{equation}
The sign of OAM of the two inputs $\Phi$ and $\Psi$ was fixed such that quantum interference could occur at the first beam splitter (see Figure~\ref{OAM_setting2}b).
While the sign of $\Phi$ only contains positive OAM, those of $\Psi$ were all negative.
Thus, $\Tilde{\Psi}$ is a $2K$ dimensional vector, whose $i$th element is the probability amplitude OAM $+i$ when $i \leq K$ and that of OAM $-i$ otherwise. 
The elements in the first half of $\Tilde{\Psi}$ were all zero because $\Psi$ contained only minus OAM. 

The output of the total system is $\Phi_{\mathrm{out}} \otimes \Psi_{\mathrm{out}}$, as shown in Figure \ref{OAM_setting2}b.
The $j$-th element of $\Phi_{\mathrm{out}}$ or $\Psi_{\mathrm{out}}$ is the probability amplitude of a photon $\Phi$ or $\Psi$ having OAM $j$ being detected at detector X when $j \leq K$. Further, it is the probability amplitude of a photon $\Phi$ or $\Psi$ having OAM $j$ being detected at detector Y when $j \geq K+1$.
Herein, the focus was placed on the cases where the two photons were detected by two different detectors.
Such probability amplitudes can be obtained by the tensor product of the latter half of $\Phi_{\mathrm{out}}$ and the first part of $\Psi_{\mathrm{out}}$ by Eqs. \eqref{phiout} and \eqref{psiout}:
\begin{equation}
    -\frac{(\alpha-\beta)^2}{2} \left[ \begin{array}{cc}
        a_1 e^{i\phi_1} \\
        \vdots \\
        a_K e^{i\phi_K} 
    \end{array} \right] \otimes \left[ \begin{array}{cc}
        b_1 e^{i\psi_1} \\
        \vdots \\
        b_K e^{i\psi_K} 
    \end{array} \right],
    \label{tens1}
\end{equation}
and by the tensor product of the latter half of $\Psi_{\mathrm{out}}$ and the first half of $\Phi_{\mathrm{out}}$:
\begin{equation}
    \frac{(\alpha + \beta)^2}{2} \left[ \begin{array}{cc}
        b_1 e^{i\psi_1} \\
        \vdots \\
        b_K e^{i\psi_K} 
    \end{array} \right] \otimes \left[ \begin{array}{cc}
        a_1 e^{i\phi_1} \\
        \vdots \\
        a_K e^{i\phi_K} 
    \end{array} \right].
    \label{tens2}
\end{equation}

\myTab[t]{Probabilities of pairs of decisions made by the asymmetric OAM system}{OAM_tab}{ll}{
 Pair of decisions & ~~Probability \\ \hline
$X:1,~~Y:1$ & \begin{tabular}{l}
$\myUnder{p}{11}= \alpha^2\beta^2 a_1^2 b_1^2 $
\end{tabular} \\ 
$X:1,~~Y:2$ & \begin{tabular}{l}
$\myUnder{p}{12}= \frac{1}{4} a_1^2b_2^2 (\alpha - \beta)^4 + \frac{1}{4} a_2^2 b_1^2 (\alpha + \beta)^4 -\frac{1}{2} a_1 a_2 b_1 b_2 (\alpha+\beta)^2 (\alpha-\beta)^2 \cos(\theta_1 - \theta_2) $
\end{tabular} \\
$X:2,~~Y:1$ & \begin{tabular}{l} $\myUnder{p}{21}=  \frac{1}{4} a_2^2 b_1^2 (\alpha - \beta)^4 + \frac{1}{4} a_1^2 b_2^2 (\alpha + \beta)^4-\frac{1}{2} a_1 a_2 b_1 b_2 (\alpha+\beta)^2 (\alpha-\beta)^2 \cos(\theta_2 - \theta_1) $ 
\end{tabular}\\
$X:2,~~Y:2$ & \begin{tabular}{l}
$\myUnder{p}{22}= \alpha^2\beta^2 a_2^2 b_2^2 $
\end{tabular} \\  
Two photons go to the same branch & \begin{tabular}{l}
$loss = 1 - \alpha^2\beta^2 +(1-4\alpha^2\beta^2)a_1a_2b_1b_2 - \frac{1}{2} (1+2\alpha^2\beta^2)(a_1^2b_2^2+a_2^2b_1^2) $
\end{tabular} \\
\hline}

Therefore, by Eqs. \eqref{tens1} and \eqref{tens2}, the probability amplitude of OAM $k_1$ is detected at X, and OAM $k_2$ is detected at Y, that is, the probability amplitude of player X choosing option $k_1$ and player Y selecting option $k_2$, is:
\begin{equation}
    \frac{1}{2} \left( (\alpha+\beta)^2 a_{k_2} b_{k_1} e^{i(\phi_{k_2}+\psi_{k_1})} - (\alpha-\beta)^2 a_{k_1} b_{k_2} e^{i(\phi_{k_1} +\psi_{k_2} )} \right).
\end{equation}
Hence, by considering the squared absolute values, the probability of player X choosing option $k_1$ and player Y selecting option $k_2$ is:
\begin{equation} 
    P(X:k_1, Y:k_2) = \frac{1}{4} a_{k_1}^2 b_{k_2}^2 (\alpha-\beta)^4 + \frac{1}{4} a_{k_2}^2 b_{k_1}^2 (\alpha + \beta)^4 - \frac{1}{2}a_{k_1}a_{k_2}b_{k_1}b_{k_2} (\alpha-\beta)^2 (\alpha + \beta)^2 \cos(\theta_{k_1}-\theta_{k_2})
    \label{OAM_prob}
\end{equation}
with $\theta_{k} := (\phi_k -\psi_k)/2$ for $k\in [K]$. Therefore, the difference between the probability of player X choosing arm $k_1$ and player Y selecting arm $k_2$ and that of player X choosing arm $k_2$ and player Y selecting arm $k_1$ is expressed as
\begin{equation}
    P~(X:k_1,Y:k_2) - P(X:k_2,Y:k_1)= 2\alpha\beta (\myUnder{a}{k_2}^2\myUnder{b}{k_1}^2 - \myUnder{a}{k_1}^2 \myUnder{b}{k_2}^2).
    \label{OAM_prob_diff}
\end{equation}
Hence, if the following condition holds true,
\begin{equation}
    \alpha\beta \ne 0,~ \myUnder{a}{k_2}\myUnder{b}{k_1} \ne \pm \myUnder{a}{k_1} \myUnder{b}{k_2} 
    \label{OAM_condi} 
\end{equation}
the difference expressed as Eq. \eqref{OAM_prob_diff} is non-zero.
Thus, $P(X:k_1, Y:k_2) \neq P(X:k_2, Y:k_1)$ is achieved; i.e., asymmetry in decision-making is realized, which is the purpose of adding the PBS in Figure \ref{OAM_setting2}.

However, conflicts can arise with a certain probability at the same time.
By substituting $k_1$ and $k_2$ of $k$, the probability of the conflict occurring with arm $k$ is expressed as:
\begin{equation} 
    P(X:k, Y:k) = \alpha^2\beta^2a_k^2 b_k^2 \neq 0.
\end{equation}


\subsection*{Results}

\begin{figure}[tb]
\centering
\includegraphics[width=\textwidth]{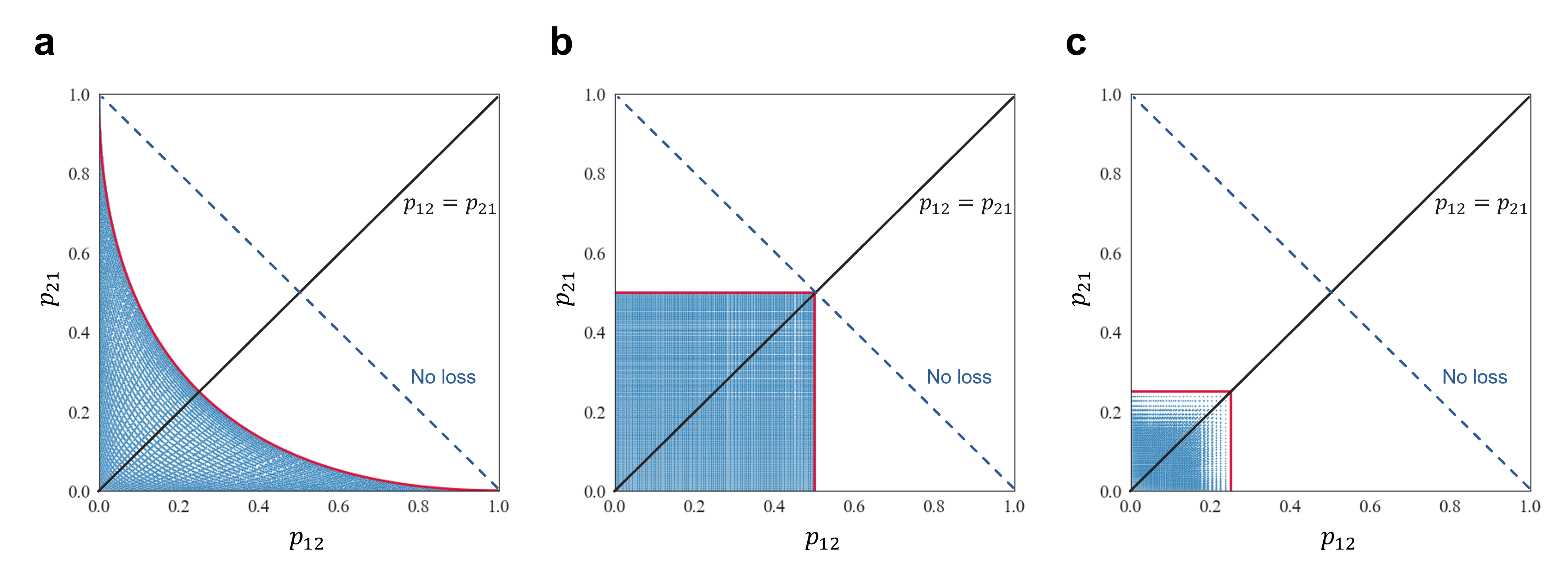}
\caption{Pairs of $(p_{12},~p_{21})$ each system is able to realize. Blue dots are the results obtained by numerical experiments. Because blue dots exist outside the $p_{12}=p_{21}$ line, asymmetric decision-making is possible by all systems. (a) Asymmetric OAM system.  (b) Entangled photon decision maker. (c) OAM attenuation. }
\label{res_pair}
\end{figure}

Next, the two-players (players X and Y), two-arms (arms 1 and 2; i.e., $K=2$) situation was examined in detail.
Table \ref{OAM_tab} summarizes the probabilities of each pair of decisions, where $p_{k_1 k_2}$ with $k_1,\ k_2\in\{1,\ 2\}$ implies that player X chooses arm $k_1$ and player Y chooses arm $k_2$.
Figure \ref{res_pair}a demonstrates the feasible pairs of $p_{12}$ and $p_{21}$ by blue-colored region on a plane, with the horizontal and vertical axes being $p_{12}$ and $p_{21}$, respectively. 
The line of $p_{12}=p_{21}$ implies the symmetric decision-making. 
As evident, the blue-colored region exists outside the $p_{12}=p_{21}$ line, thus validating the feasibility of asymmetric decision-making.

However, the asymmetric OAM system cannot realize all combinations of $(p_{12},p_{21})$. 
For example, $(p_{12}, p_{21})=(0.5, 0.5)$ is outside the feasible zone. 
Indeed, the red curve in Figure \ref{res_pair}a shows the boundary between the feasible and infeasible zones of $(p_{12},p_{21})$. The first right side of the boundary belongs to the impossible zone, whereas the lower left side 
belongs to the possible zone. 
This boundary also corresponds to the cases without loss.
The formula of this boundary is expressed as:
\begin{equation}
    2(p_{12} + p_{21}) = 1+ (p_{12} - p_{21})^2.
    \label{OAM_bound}
\end{equation}
See the Supplementary Information for the derivation of Eq. \eqref{OAM_bound}.

\begin{figure}[tb]
\centering
\includegraphics[width=\textwidth]{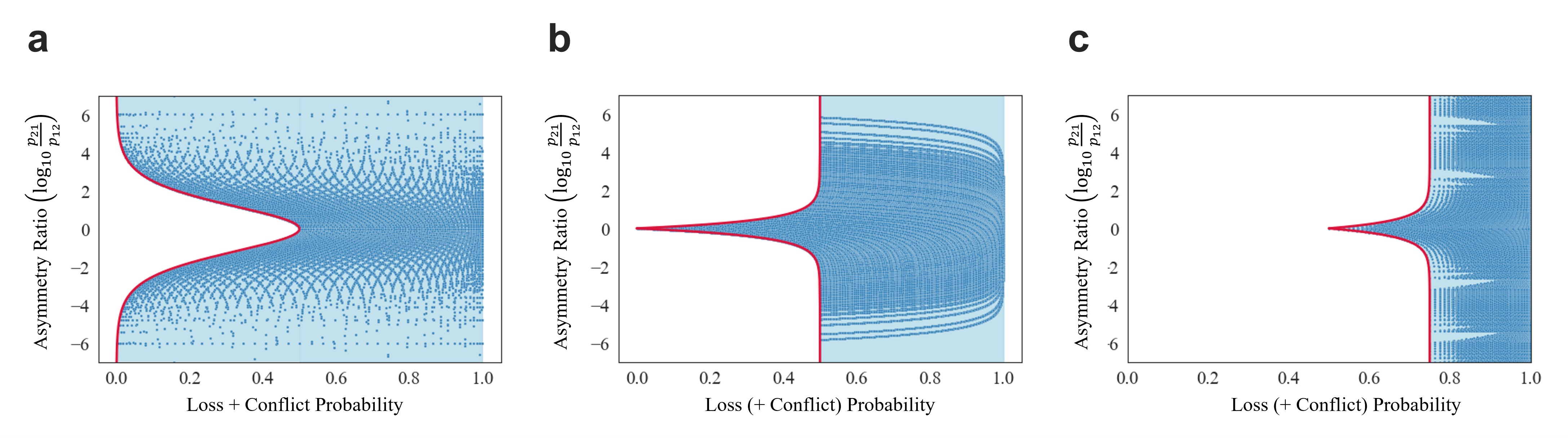}
\caption{The relationship between the Loss Probability plus Conflict Probability and the Asymmetry Ratio of each system. Blue dots are the results obtained by numerical experiments. The light blue area is the area mathematical consideration can prove that each system can realize. (a) Asymmetric OAM system. (b) Entangled photon decision maker. (c) OAM attenuation.}
\label{res_ratio}
\end{figure}

Thus, the \textit{conflict probability}, the probability of both players choosing the same arm, is defined as $p_{12} + p_{21}$, the \textit{asymmetry ratio} of the decision-making as $p_{21}/p_{12}$, and the \textit{loss probability} of photons as $1-(p_{11}+p_{12}+p_{21}+p_{22})$.
Figure \ref{res_ratio}a shows the relationship between the conflict probability plus loss probability and symmetry ratio.
The red-colored boundary in Figure \ref{res_ratio}a denotes the minimum-loss-plus-conflict boundary.
By defining the conflict probability plus loss probability as $x$ and the asymmetry ratio as $y$, the formula is expressed as:
\begin{equation}
\label{eq:yofxOAM}
    y=\begin{cases}
        \cfrac{(1+\sqrt{1-2x})^2}{(1-\sqrt{1-2x})^2} ~~& \text{when } y \ge 1 ,\\
        \\  
        \cfrac{(1-\sqrt{1-2x})^2}{(1+\sqrt{1-2x})^2}~~& \text{when } y \le 1.
    \end{cases}
\end{equation}
The detailed derivation of Eq. \eqref{eq:yofxOAM} is presented in Supplementary Information.  

In the entangled photon decision maker, described later, 50\% loss or conflict is inevitable in obtaining any asymmetry ratio.
This rate is smaller than the smallest percentage necessary to realize all asymmetry ratios in the OAM attenuation.
For situations when a lower rate of loss or conflict is appealing, an extreme asymmetry ratio, such as more than $100$ or smaller than $0.01$, is obtained by the asymmetry OAM system.
Therefore, the decision-making system using OAM is more suitable when inequality between players is serious such that more powerful affirmative actions are necessary.

\subsection*{Obtaining a specific asymmetry ratio}

In terms of application, the method to obtain the intended asymmetry ratio must be determined. 
First, any asymmetry ratio is possible while avoiding decision conflicts.
Based on the results presented in Table \ref{OAM_tab}, the conflict probability becomes zero when $a_2=b_1=0$ or $a_1=b_2=0$.
Note that the loss probability is not zero.

\begin{figure}[t]
\centering
\includegraphics[width=0.8\textwidth]{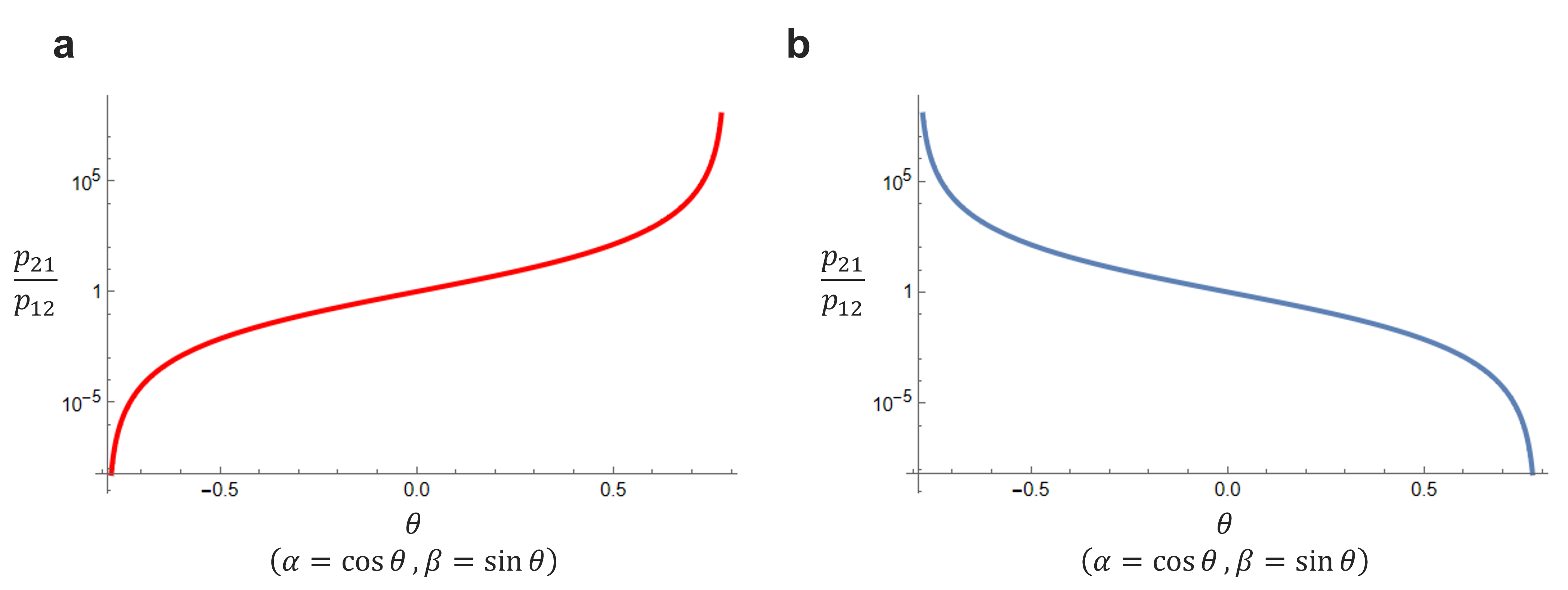}
\caption{(a) The relationship between $\theta$ and $r$ when $a_1=b_2=0$. (b) The relationship between $\theta$ and $r$ when $a_2=b_1=0$. In both cases, any asymmetry ratio can be achieved by $\theta$, $-\pi/4 \leq \theta \leq \pi/4$.}
\label{OAM_make_ratio}
\end{figure}

Let $r$ be the desired asymmetry ratio.
When $a_2=b_1=0$, the asymmetry ratio is expressed as:
\begin{equation} 
\label{eq:r}
    r=\frac{p_{21}}{p_{12}} = \frac{(\alpha+\beta)^4}{(\alpha-\beta)^4}.
\end{equation}
By introducing $\theta$ such that $\alpha =\cos\theta, \beta = \sin\theta$, 
Eq. \eqref{eq:r} becomes
\begin{equation} 
    r=\frac{(\cos\theta+\sin\theta)^4}{(\cos\theta-\sin\theta)^4} .
    \label{OAM_eq_ratio}
\end{equation}
Organizing Eq. \eqref{OAM_eq_ratio} about $\theta$, we obtain
\begin{equation} 
    3-3r-\cos 4\theta + r\cos4\theta + 4\sin 2\theta + 4r\sin2\theta = 0.
    \label{OAM_eq_ratio2}
\end{equation}
By solving Eq. \eqref{OAM_eq_ratio2}, we obtain $\alpha$ and $\beta$ to realize $r$ without conflicts.
Figure \ref{OAM_make_ratio}a shows the relationship 
between $\theta$ and $r$ 
based on Eq. \eqref{OAM_eq_ratio}, 
showing that $r$ can take every value with $\theta$ from $-\pi/4$ to $\pi/4$.
The realization of any $r$ is significant because the degree of asymmetry can be balanced depending on the current inequality between players.

Similarly, when $a_1=b_2=0$, the asymmetry ratio is:
\begin{equation} 
    r=\frac{p_{21}}{p_{12}} = \frac{(\alpha-\beta)^4}{(\alpha+\beta)^4} 
\end{equation}
which is reformulated using $\theta$ as
\begin{equation} 
    r=\frac{(\cos\theta-\sin\theta)^4}{(\cos\theta+\sin\theta)^4}.
\end{equation}
Hence, 
by solving the following Eq. \eqref{OAM_eq_ratio3}, we obtain $\alpha$ and $\beta$ to realize $r$ without conflicts.
\begin{equation}
    3-3r-\cos 4\theta + r\cos4\theta - 4\sin 2\theta - 4r\sin2\theta = 0 \label{OAM_eq_ratio3}
\end{equation}
Figure \ref{OAM_make_ratio}b shows the relationship between $\theta$ and $r$ based on Eq. \eqref{OAM_eq_ratio3}, showing that $r$ can acquire every value with $\theta$ from $0$ to $\pi/4$.

\myTab[b]{Probabilities of decisions when BS is added to symmetric OAM system instead of PBS}{BS_tab}{ll}{
 Decisions & ~~Probability \\ \hline
$X:1,~~Y:1$ & \begin{tabular}{l} $\myUnder{p}{11}= \frac{1}{4} a_1^2 b_1^2 $ \end{tabular} \\ 
$X:1,~~Y:2$ & \begin{tabular}{l} $\myUnder{p}{12}= a_2^2 b_1^2 $ \end{tabular} \\
$X:2,~~Y:1$ & \begin{tabular}{l} $\myUnder{p}{21}=  a_1^2 b_2^2  $  \end{tabular}\\
$X:2,~~Y:2$ & \begin{tabular}{l} $\myUnder{p}{22}= \frac{1}{4} a_2^2 b_2^2 $ \end{tabular} \\
Two photons go to the same branch & \begin{tabular}{l} $loss= \frac{3}{4} (1-a_2^2 b_1^2-a_1^2 b_2^2) $ \end{tabular} \\\hline}

\subsection*{Origin of the asymmetry}

The difference between the asymmetric and symmetric OAM system is the existence of the PBS in the system and the addition of polarization to the photon state.
The polarization of the photon state is expressed by two parameters: $\alpha$ and $\beta$.
With the probabilities of $|\alpha|^2$ and $|\beta|^2$, photons are detected as horizontal and vertical polarizations, respectively.
These parameters satisfy $|\alpha|^2+|\beta|^2=1$.
When $\alpha=0$ or $\beta=0$, the photons simply transmit or are reflected at the PBS.
Therefore, quantum interference does not occur at the PBS, with PBS playing no role; this situation corresponds to the symmetric OAM system \cite{amakasu}.
However, when $\alpha \neq 0$ and $\beta \neq 0$, whether the photons are transmitted or reflected at the PBS is decided stochastically.
Therefore, quantum interference can occur.
Thus, the occurrence of quantum interference at the PBS renders a difference between the asymmetric and symmetric OAM systems.

Indeed, asymmetric decision-making is possible via the addition of both PBS and BS.
Table \ref{BS_tab} lists the probabilities of pairs of decisions in the case where BS is added instead of PBS.
When BS was added instead of PBS, fewer states could be achieved.
For example, by adding PBS to the symmetric OAM system, any nonnegative asymmetry ratio can be achieved without conflicts.
This is because parameters $\alpha$ and $\beta$ possess the degree of freedom even if $a_1,~~a_2,~~b_1~~b_2$ are set to $(0,~~1,~~1,~~0)$ or $(1,~~0,~~0,~~1)$ to render conflict probability zero.
However, when attempting to render conflicts free in the system where BS is added, only two states can be realized: $p_{12}=0, ~~p_{21}=1$ or $p_{12}=1, ~~p_{21}=0$.
Therefore, the addition of PBS to the symmetric OAM system yields superior results.

\section*{Asymmetric decision-making by entangled photon decision maker}

\begin{figure}[t]
    \centering
    \includegraphics[width=\textwidth]{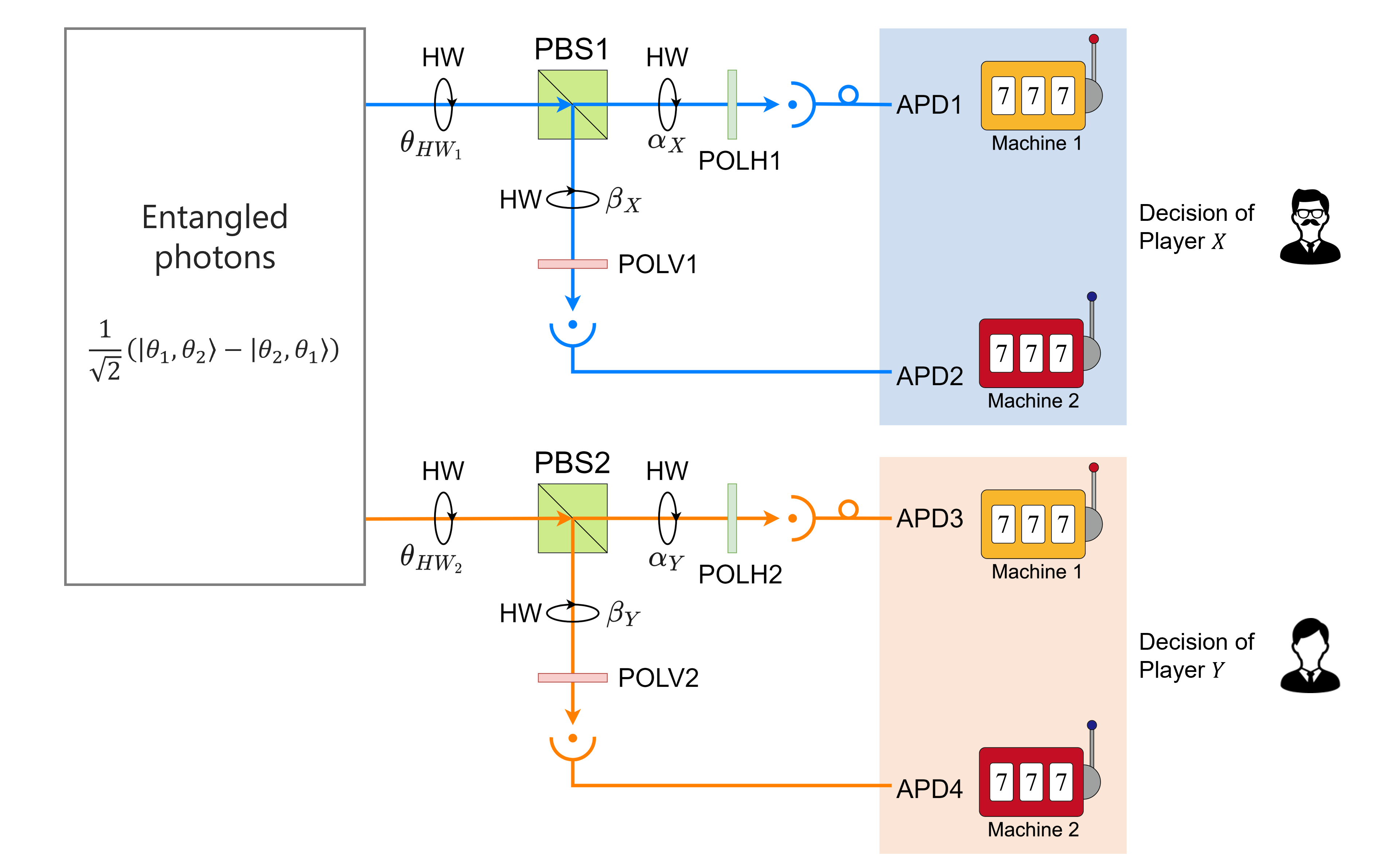}
    \caption{Experimental setup of the entangled photon decision maker. PBS: Polarization Beam Splitter, HW: Half-wave Plate, APD: Avalanche Photodiode, POLH: Polarizer (allowing only horizontal polarization to pass), POLV: Polarizer (allowing only vertical polarization to pass)}
    \label{etg_setting}
\end{figure}

This section presents the entangled photon decision-maker that can fulfill asymmetric decision-making, particularly for the two-players, two-arms bandit problem. 
Figure \ref{etg_setting} shows a schematic of the entangled decision-maker.
The input to the system is two entangled photons.
One photon entering PBS 1 decides player X's choice while another entering PBS 2 decides player Y's.
In a previous study [\citeonline{chauvet}], conflict-free, symmetric decision-making among two players was theoretically and experimentally demonstrated. 
The system shown in Figure \ref{etg_setting} realizes the asymmetry by discarding photons with specific probabilities at the polarizers before APDs or avalanche photodiodes.
This system is different from that in the previous study \cite{chauvet} owing to the presence of polarizers.
Note that herein, a specific input is assumed:
\begin{equation}
    \frac{1}{\sqrt{2}}\Bigl(\ket{\theta_1, \theta_2} - \ket{\theta_2, \theta_1} \Bigr).
    \label{etg_input}
\end{equation}
This is a superposition of the following two states.
One is the state with photons with polarizations of $\theta_1$ and $\theta_2$ entering the PBS 1 and 2, respectively.
The other is the state with photons with polarization $\theta_2$ and $\theta_1$ entering the PBS 1 and 2, respectively.
In particular, the latter state employs a $\pi$ phase shift to consider the minus sign of the second term in Eq. \eqref{etg_input}, i.e. the second term is actually $e^{i\pi} \ket{\theta_2, \theta_1}$.
Herein, the polarizations of $\theta_1$ and $\theta_2$ are orthogonal to each other and satisfy the following condition.
\begin{equation}
    \theta_2 = \theta_1 + \frac{\pi}{2}.
    \label{etg_cond}
\end{equation}

\subsection*{Formulation}
Next, the probabilities corresponding to pairs of decisions were derived.
Here, we give the space to represent the photon states and the corresponding decision-making.
The option 1 and 2 are represented by vectors $\ket{1} = [~1,~0~]^\top,~~\ket{2} = [~0,~1~]^\top$, respectively.
By associating options with detectors, photon states can be described in the Hilbert space:
\begin{equation}
    \mathcal{H} = \mathrm{span}\{ \ket{1},\ket{2} \} \simeq \mathbb{C}^2.
\end{equation}
The first photon, present in the upper part of the system, corresponds to options 1 and 2 when detected at APD 1 and APD 2, respectively.
Similarly, the output of the second photon, present in the lower part, corresponds to options 1 and 2 when detected at APD 3 and APD 4, respectively.
Therefore, the output of the total system is in the Hilbert space $\mathcal{H} \otimes \mathcal{H}$.

First, consider the first term of Eq. \eqref{etg_input}, $\ket{\theta_1, ~~\theta_2}$.
The output of the photon injected into PBS 1 is:
\begin{equation}
    \phi_1'=
    \left[ \begin{array}{l}
         \cos\alpha_X \cos(2\theta_{\mathrm{HW}_1} - \theta_1) \\
         \sin\beta_X \sin(2\theta_{\mathrm{HW}_1} - \theta_1)
    \end{array} \right].
\end{equation}
The first element is the probability amplitude of the photon detected at APD1 or the horizontal component, 
whereas the second one is that of the photon detected at APD2 or the vertical component.
Similarly, the output of the photon entering PBS 2 is expressed as:
\begin{equation}
    \phi_1''=
    \left[ \begin{array}{l}
         \cos\alpha_Y \cos(2\theta_{\mathrm{HW}_2} - \theta_2) \\
         \sin\beta_Y \sin(2\theta_{\mathrm{HW}_2} - \theta_2)
    \end{array} \right].
\end{equation}
Here, $\alpha_X, \alpha_Y, \beta_X, \beta_Y$ are the orientations of the polarizers. 
The first element implies the probability amplitude of the photon detected at APD3, whereas the second one is that of the photon detected at APD4.
Therefore, by considering the tensor of $\phi_1'$ and $\phi_1''$, the output of the first term of Eq. \eqref{etg_input}, $\ket{\theta_1,~~\theta_2}$, is expressed as:
\begin{equation}
    \phi_1=\phi_1' \otimes \phi_1'' = 
    \left[ \begin{array}{l}
         \cos\alpha_X \cos(2\theta_{\mathrm{HW}_1} - \theta_1) \cos\alpha_Y \cos(2\theta_{\mathrm{HW}_2} - \theta_2) \\
         \cos\alpha_X \cos(2\theta_{\mathrm{HW}_1} - \theta_1) \sin\beta_Y \sin(2\theta_{\mathrm{HW}_2} - \theta_2) \\
         \sin\beta_X \sin(2\theta_{\mathrm{HW}_1} - \theta_1) \cos\alpha_Y \cos(2\theta_{\mathrm{HW}_2} - \theta_2) \\
         \sin\beta_X \sin(2\theta_{\mathrm{HW}_1} - \theta_1) \sin\beta_Y \sin(2\theta_{\mathrm{HW}_2} - \theta_2)
    \end{array} \right].
\end{equation}
The first element is the probability amplitude of two photons being detected at APD1 and APD3, whereas the second is that of them being detected at APD1 and APD4, the third is that of them being detected at APD2 and APD3, and the fourth is that of them being detected at APD2 and APD4.

Next, consider the second term of Eq. \eqref{etg_input}, $\ket {\theta_2, ~~\theta_1}$.
Similar to the first term, the output of the second term is expressed as:
\begin{equation}
    \phi_2=
    \left[ \begin{array}{l}
         \cos\alpha_X \cos(2\theta_{\mathrm{HW}_1} - \theta_2) \cos\alpha_Y \cos(2\theta_{\mathrm{HW}_2} - \theta_1) \\
         \cos\alpha_X \cos(2\theta_{\mathrm{HW}_1} - \theta_2) \sin\beta_Y \sin(2\theta_{\mathrm{HW}_2} - \theta_1) \\
         \sin\beta_X \sin(2\theta_{\mathrm{HW}_1} - \theta_2) \cos\alpha_Y \cos(2\theta_{\mathrm{HW}_2} - \theta_1) \\
         \sin\beta_X \sin(2\theta_{\mathrm{HW}_1} - \theta_2) \sin\beta_Y \sin(2\theta_{\mathrm{HW}_2} - \theta_1)
    \end{array} \right].
\end{equation}
Considering the superposition of $\phi_1$ and $\phi_2$ with the factor of $1/\sqrt{2}$ as expressed in Eq. \eqref{etg_input}, the output probability amplitudes is expressed as:
\begin{equation}
    \frac{1}{\sqrt{2}}\left(\phi_1 - \phi_2\right) = \frac{1}{\sqrt{2}}
    \left[ \begin{array}{l}
         \cos\alpha_X \cos(2\theta_{\mathrm{HW}_1} - \theta_1) \cos\alpha_Y \cos(2\theta_{\mathrm{HW}_2} - \theta_2) - \cos\alpha_X \cos(2\theta_{\mathrm{HW}_1} - \theta_2) \cos\alpha_Y \cos(2\theta_{\mathrm{HW}_2} - \theta_1) \\
         \cos\alpha_X \cos(2\theta_{\mathrm{HW}_1} - \theta_1) \sin\beta_Y \sin(2\theta_{\mathrm{HW}_2} - \theta_2) - \cos\alpha_X \cos(2\theta_{\mathrm{HW}_1} - \theta_2) \sin\beta_Y \sin(2\theta_{\mathrm{HW}_2} - \theta_1) \\
         \sin\beta_X \sin(2\theta_{\mathrm{HW}_1} - \theta_1) \cos\alpha_Y \cos(2\theta_{\mathrm{HW}_2} - \theta_2) - \sin\beta_X \sin(2\theta_{\mathrm{HW}_1} - \theta_2) \cos\alpha_Y \cos(2\theta_{\mathrm{HW}_2} - \theta_1) \\
         \sin\beta_X \sin(2\theta_{\mathrm{HW}_1} - \theta_1) \sin\beta_Y \sin(2\theta_{\mathrm{HW}_2} - \theta_2) - \sin\beta_X \sin(2\theta_{\mathrm{HW}_1} - \theta_2) \sin\beta_Y \sin(2\theta_{\mathrm{HW}_2} - \theta_1)
    \end{array} \right].
\end{equation}
Hence, by considering the squared absolute value of each element and the condition of Eq. \eqref{etg_cond} along with assuming that $\theta_{\mathrm{HW}_1}$ is equal to $\theta_{\mathrm{HW}_2}$, the probabilities are expressed as:
\begin{equation}
    \left|\frac{1}{\sqrt{2}}\left(\phi_1 - \phi_2\right)\right|^2=
    \frac{1}{2} \left[\begin{array}{c}
        0 \\
        \sin^2\beta_Y \cos^2\alpha_X \\
         \sin^2\beta_X \cos^2\alpha_Y \\
        0
    \end{array}\right].
    \label{etg_probs}
\end{equation}

Herein, the first element is the probability of photons detected at APD1 and APD3, essentially implying that both players X and Y choose arm 1. Further, the fourth element is the probability of photon detection at APD2 and APD4, corresponding to both players X and Y selecting arm 2. Owing to the entanglements, the corresponding probabilities are zero ($p_{11}=p_{22}=0$).

The second term implies that photons are received by APD1 and APD4, corresponding to player X choosing arm 1 and player Y choosing arm 2. Similarly, the third term implies that photons are received by APD2 and APD4, thus indicating that players X and Y select arms 2 and 1, respectively.

The original entangled-photon decision-maker is with $\alpha_i$ being $0$ and $\beta_i$ being $\pi/2$, which results in the second and the third term being $1/2$. Consequently, perfect equality is ensured. With the inclusion of polarizers, such equality can be broken, as expressed in Eq. \eqref{etg_probs}.

\subsection*{Results}

\myTab[t]{Probabilities of decisions in the entangled-photon decision maker}{etg_tab}{llll}{
Photon1 & Photon2 & Decisions & Probability \\ \hline
APD1 & APD3 & $X:1,~~Y:1$ & $\myUnder{p}{11}=0$ \\
APD1 & APD4 & $X:1,~~Y:2$ & $\myUnder{p}{12}=  \sin{\beta_Y}^2 \cos{\alpha_X}^2/2 $ \\
APD2 & APD3 & $X:2,~~Y:1$ & $\myUnder{p}{21}=  \sin{\beta_X}^2 \cos{\alpha_Y}^2/2 $ \\
APD2 & APD4 & $X:2,~~Y:2$ & $\myUnder{p}{22}=0$ \\ \hline
}

Table \ref{etg_tab} summarizes the probabilities corresponding to pairs of decisions derived using Eq. \eqref{etg_probs}.
Note that conflicts never happen in this system; this implies that two players always choose different options, whereas asymmetric decision-making ($p_{12} \neq p_{21}$) can be achievable by adequately setting $\alpha_i$ and $\beta_i$ ($i=1,2$).
Nonetheless, $p_{12}$ and $p_{21}$ cannot be larger than $0.5$; thus, photon losses are inevitable unless $\alpha_X=\alpha_Y=0$ and $\beta_X=\beta_Y=\pi/2$.

The blue-colored region in Figure \ref{res_pair}b presents the achievable pairs of $p_{12}$ and $p_{21}$ in the diagram of $p_{12}$ and $p_{21}$ in the horizontal and vertical axes, respectively.
This demonstrates clearly that any $p_{12}$ and $p_{21}$ are accepted if they are equal or less than $0.5$; thus, the blue-colored feasible zone resulted in a square area in the lower left corner of the diagram. 

Figure \ref{res_ratio}b shows the relationship between the conflict probability plus loss probability and asymmetry ratio.
The mathematical formula of the red-lines border is expressed as:
\begin{equation}
    y=\begin{cases}
    \cfrac{1}{1-2x} ~~& \text{when } y \ge 1, \\
    \\  
    1-2x~~& \text{when } y \le 1,
    \end{cases}
\end{equation}
The detailed derivation is presented in the Supplementary Information. 
Figure \ref{res_ratio}b shows that at most, 50\% loss of photons must be tolerated to obtain any asymmetry ratios.
The loss is minimal for the symmetric case where $p_{12}=p_{21}=0.5$.
This implies that the entangled photon decision maker is appropriate when maintaining an even treatment is ideal because the two players are almost equal.

\subsection*{Obtaining a specific asymmetry ratio}
The intended asymmetry ratio $r$ must be obtained by the entanglement system.
For this, focus was placed on two cases: $p_{12}=0.5$ and $p_{21}=0.5$, because the loss probability is the smallest on the possible zone in Figure \ref{res_pair}b under the same asymmetry ratio.
When $p_{12}=0.5$, 
\begin{equation}
    r=\frac{p_{21}}{p_{12}}=\sin^2 \alpha_X \cos^2 \alpha_Y \leq 1.   
    \label{etg_eq1}
\end{equation}
When $p_{21}=0.5$,
\begin{equation}
    r=\frac{p_{21}}{p_{12}}=\frac{1}{\sin^2 \alpha_Y \cos^2 \alpha_X} \geq 1.    
    \label{etg_eq2}
\end{equation}
Therefore, for an asymmetry ratio greater than $1$, Eq. \eqref{etg_eq2} must be solved.
Otherwise, Eq. \eqref{etg_eq1} must be solved.
Thus, the parameters required to achieve the intended $r$ can be obtained.
However, loss becomes great when the desired ratio is extreme because either $p_{12}$ or $p_{21}$ is fixed to $0.5$.

\section*{Asymmetric decision-making by OAM attenuation}
\begin{figure}[t]
\centering
\includegraphics[width=\textwidth]{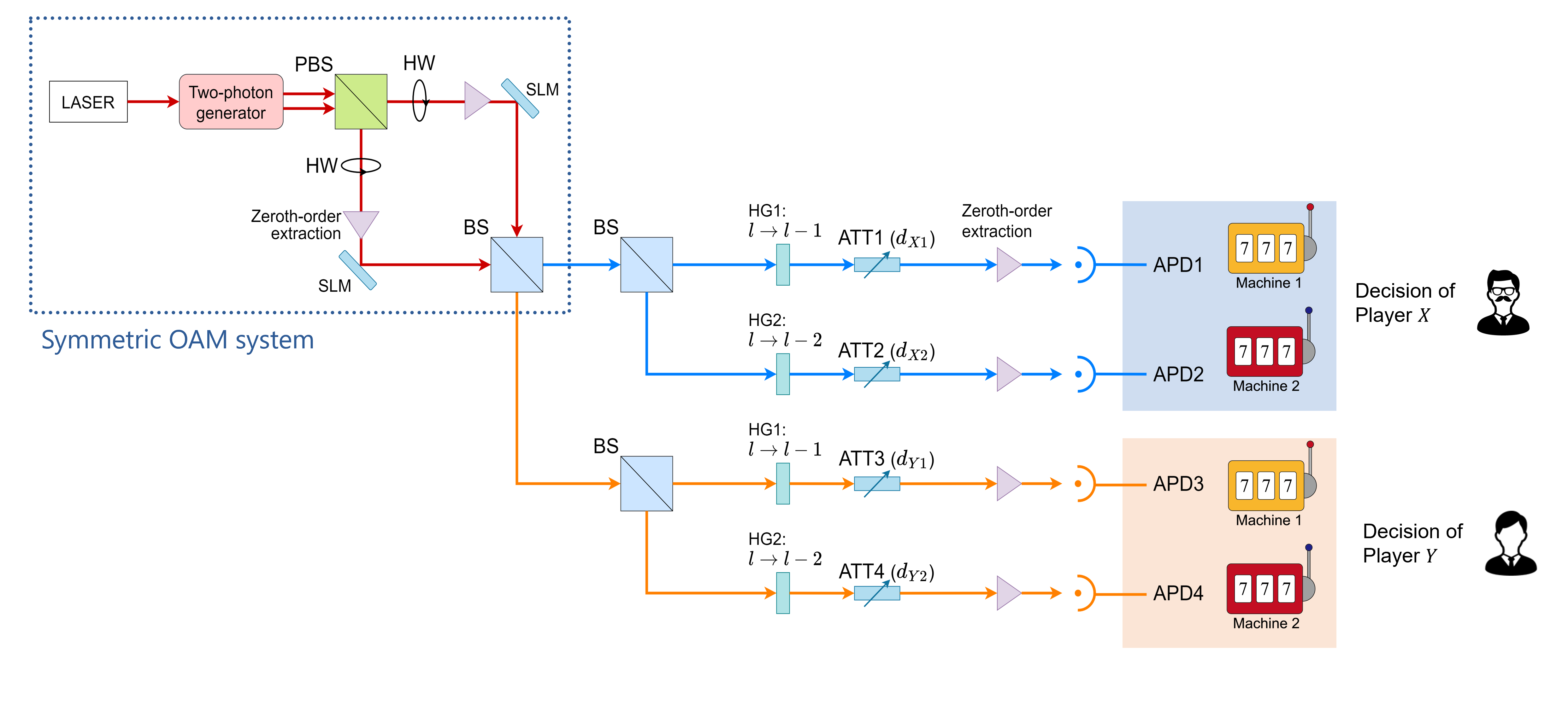}
\caption{Decision-making system by OAM attenuation. HG= Hologram, ATT=Attenuator, BS=Beam Splitter, SLM=Spatial Light Modulator.}
\label{att_setting}
\end{figure}

\subsection*{Formulation}

This section presents decision-making by the OAM attenuation system.
Figure \ref{att_setting} shows the architecture of the OAM attenuation setup.
The two upper detectors, APD1 and APD2, are related to the decision-making of player X, whereas the two lower ones, APD3 and APD4, are related to the decision-making of player Y.
The probability of each pair of decisions is the probability of the symmetric decision-making system multiplied by attenuation rates.
Thus, the probability of player X choosing option $i$ and player Y selecting option $j$ is expressed as:
\begin{equation}
    p_{ij} = \left( a_i ^2 b_j ^2 + a_j^2 b_i^2 -2 a_i a_j b_i b_j \cos (\theta_i - \theta_j) \right) d_{X i}^2 d_{Yj}^2, ~~i, j \in \{1, 2\}.
\end{equation}
When $d_{X1} \neq d_{X2}$ and $d_{Y1} \neq d_{Y2}$, the OAM attenuation system fulfills asymmetric decision-making. 

\subsection*{Results}

\myTab[t]{Probabilities of decisions in the OAM attenuation}{att_tab}{llll}{
Photon1 & Photon2 & Decisions & Probability \\ \hline
APD1 & APD3 & $X:1,~~Y:1$ & $\myUnder{p}{11}=0$ \\
APD1 & APD4 & $X:1,~~Y:2$ & $\myUnder{p}{12}= \frac{1}{4} \left( a_1^2 b_2^2 + a_2^2 b_1^2 - 2 a_1 a_2 b_1 b_2 \cos(\theta_1-\theta_2) \right) d_{X1}^2 d_{Y2}^2 $ \\
APD2 & APD3 & $X:2,~~Y:1$ & $\myUnder{p}{21}= \frac{1}{4} \left( a_2^2 b_1^2 + a_1^2 b_2^2 - 2 a_1 a_2 b_1 b_2 \cos(\theta_1-\theta_2) \right) d_{X2}^2 d_{Y1}^2 $ \\
APD2 & APD4 & $X:2,~~Y:2$ & $\myUnder{p}{22}=0$ \\ \hline
}

Figure \ref{res_pair}c shows the available pairs of $p_{12}$ and $p_{21}$.
As evident, photon loss is inevitable in this system.
Figure \ref{res_ratio}c shows the relationship between the loss probability and asymmetry ratio.
If 75\% of photon loss is allowed, all asymmetry ratios can be obtained.
However, 75\% is the highest probability for achieving all asymmetry ratios among the three systems.
Moreover, the OAM attenuation system is without any conflicts because it is based on the symmetric OAM system.
The mathematical formula of the red-lined boundary in Figure \ref{res_ratio}c is expressed as:
\begin{equation}
\label{eq:yofxOAMattenuation}
    y = \begin{cases} 3 - 4x ~~ \text{when} ~~ y \leq 1, \\
        \cfrac{1}{3-4x} ~~ \text{when} ~~ y\geq 1 .   \end{cases}
\end{equation}
The derivation of Eq. \eqref{eq:yofxOAMattenuation} is presented in the Supplementary Information.

\subsection*{Obtaining a specific asymmetry ratio}
Next, we explore how an intended asymmetry ratio $r$ can be obtained in the OAM attenuation setup.
For this, we focus on the case $p_{12}=0.25$ or $p_{21}=0.25$, because the loss probability is the smallest within the possible zone in Figure \ref{res_pair}c for this particular asymmetry ratio.
When $p_{12} = 0.25$, 
\begin{equation}
    r= \frac{p_{21}}{p_{12}}=\left( a_2^2 b_1^2 + a_1^2 b_2^2 - 2 a_1 a_2 b_1 b_2 \cos(\theta_1-\theta_2) \right) d_{X2}^2 d_{Y1}^2 \leq 1.   
    \label{att_eq1}
\end{equation}
When $p_{21} = 0.25$,
\begin{equation}
    r= \frac{1}{\left( a_2^2 b_1^2 + a_1^2 b_2^2 - 2 a_1 a_2 b_1 b_2 \cos(\theta_1-\theta_2) \right) d_{X1}^2 d_{Y2}^2} \geq 1.   
    \label{att_eq2}
\end{equation}
If we introduce $u,~v$ such that $a_1 = \cos{u},~~a_2 = \sin{u}$ and $b_1 = \cos{v},~~b_2=\sin{v}$, then Eq. \eqref{att_eq1} becomes:
\begin{equation}
    r = \frac{p_{21}}{p_{12}} = \frac{1}{2} \left( 1- \cos{2u} \cos{2v} - \cos(\theta_1 - \theta_2) \sin{2u} \sin{2v} \right)  d_{X2}^2 d_{Y1}^2,
    \label{att_eq3}
\end{equation}
and Eq. \eqref{att_eq2} becomes:
\begin{equation}
    r = \frac{2}{\left(1- \cos{2u} \cos{2v} - \cos(\theta_1 - \theta_2) \sin{2u} \sin{2v} \right)  d_{X1}^2 d_{Y2}^2} .
    \label{att_eq4}
\end{equation}
When the intended $r$ is greater than $1$, the parameters required to achieve $r$ can be obtained by solving Eq. \eqref{att_eq3} about $u$ and $v$.
Otherwise, Eq. \eqref{att_eq4} must be solved.
However, loss becomes great when the desired ratio is extreme because either $p_{12}$ or $p_{21}$ is fixed to $0.25$.

\section*{Conclusion}
This study analysed the asymmetric collective decision-making via quantum properties, quantum interference, and entanglement to explore the possibility of how to implement affirmative actions to reduce disparities in the case of two players.
Asymmetry in decision-making was successfully realized by three systems: the asymmetric OAM system, asymmetric entangled photon decision-maker, and OAM attenuation. 
The probability of players X and Y choosing options 1 and 2, respectively, which is denoted by $p_{12}$, can be different from the probability of players X and Y choosing options 2 and 1, respectively, which is denoted by $p_{21}$. Thus, $p_{12} \neq p_{21}$ was demonstrated to be achievable. 
Previous studies were limited to symmetric collective decision-making ($p_{12} = p_{21}$).
Through asymmetric joint decisions, the unfairness and inequalities among agents, which might be innate prior to the games, could be modulated.

Further, herein, several limitations in achieving the asymmetry were clarifed.
In all systems, some impossible pairs of $p_{12}$ and $p_{21}$ were shown to exist. 
Related, loss of photons or decision conflicts were inevitably presented with a certain probability.
For example, only two combinations of $(p_{12}, ~p_{21})$ were achievable without any loss or conflicts in the asymmetric OAM system, whereas the asymmetric entangled photon decision maker could realize zero-loss decisions only in the symmetric case.
However, the OAM attenuation setup could not realize any situation without loss or conflicts.
The minimum probability of loss or conflicts to achieve any ratio of $p_{12}$ and $p_{21}$ was the smallest, that is, 50\%, in the asymmetric OAM system and asymmetric entangled photon decision maker.
In case of the OAM attenuation, 75\% loss was required to be tolerated to obtain any ratio. 
This study analytically clarified the boundary between the feasible and infeasible zones of the combination of $(p_{
12}, ~p_{21})$ for each of the three systems.

However, the existence of unavailable pairs of $p_{12}$ and $p_{21}$ may not be a serious concern because any ratio of $p_{12}$ and $p_{21}$ can be accomplished by accepting the certain probability of loss or conflicts.
Further, to enable affirmative actions tackle inequalities, all pairs were not required if an appropriate degree of asymmetry could be achieved.
The formula for configuring the physical parameters in the quantum systems to realize a specified value of the ratio of $p_{12}$ and $p_{21}$ was analytically derived. 
Finally, because all systems were tuned for only two-player situations, extensions to cases with more players are expected.

In situations when powerful affirmative action is needed as the inequality between players is serious, the asymmetric OAM system is best suited because it exhibits superior performance with low conflict rates and photon loss.
Whereas the use of the asymmetric entangled photon decision maker is advisable owing to a minor loss of photons and lower conflict rate when the inequality is rather small and moderate affirmative action is sufficient.
This study contributes to extending the photonic and quantum collective decision-making to asymmetric properties, thus paving the way toward covering a broader sense of equality and social welfare based on quantum principles.

\section*{Acknowledgments}
This work was supported in part by the CREST project (JPMJCR17N2) funded by the Japan Science and Technology Agency, Grants-in-Aid for Scientific Research (JP20H00233) and Transformative Research (JP22H05197) funded by the Japan Society for the Promotion of Science (JSPS), and CNRS-UTokyo Excellence Science Joint Research Program. AR was funded by JSPS as an International Research Fellow.

\bibliography{bibliography}

\begin{thebibliography}{10}
\urlstyle{rm}
\expandafter\ifx\csname url\endcsname\relax
  \def\url#1{\texttt{#1}}\fi
\expandafter\ifx\csname urlprefix\endcsname\relax\def\urlprefix{URL }\fi
\expandafter\ifx\csname doiprefix\endcsname\relax\def\doiprefix{DOI: }\fi
\providecommand{\bibinfo}[2]{#2}
\providecommand{\eprint}[2][]{\url{#2}}

\bibitem{daw2006cortical}
\bibinfo{author}{Daw, N.~D.}, \bibinfo{author}{O'doherty, J.~P.},
  \bibinfo{author}{Dayan, P.}, \bibinfo{author}{Seymour, B.} \&
  \bibinfo{author}{Dolan, R.~J.}
\newblock \bibinfo{journal}{\bibinfo{title}{Cortical substrates for
  explovilloresi, p.ratory decisions in humans}}.
\newblock {\emph{\JournalTitle{Nature}}} \textbf{\bibinfo{volume}{441}},
  \bibinfo{pages}{876--879} (\bibinfo{year}{2006}).

\bibitem{sutton2018reinforcement}
\bibinfo{author}{Sutton, R.~S.} \& \bibinfo{author}{Barto, A.~G.}
\newblock \emph{\bibinfo{title}{Reinforcement learning: An introduction}}
  (\bibinfo{publisher}{MIT press}, \bibinfo{year}{2018}).

\bibitem{Auer}
\bibinfo{author}{Auer, P.}, \bibinfo{author}{Cesa-Bianchi, N.} \&
  \bibinfo{author}{Fischer, P.}
\newblock \bibinfo{journal}{\bibinfo{title}{Finite-time analysis of the
  multiarmed bandit problem}}.
\newblock {\emph{\JournalTitle{Machine learning}}}
  \textbf{\bibinfo{volume}{47(2)}}, \bibinfo{pages}{235--256}
  (\bibinfo{year}{2002}).

\bibitem{march1991exploration}
\bibinfo{author}{March, J.~G.}
\newblock \bibinfo{journal}{\bibinfo{title}{Exploration and exploitation in
  organizational learning}}.
\newblock {\emph{\JournalTitle{Organization science}}}
  \textbf{\bibinfo{volume}{2}}, \bibinfo{pages}{71--87} (\bibinfo{year}{1991}).

\bibitem{chauvet}
\bibinfo{author}{Chauvet, N.} \emph{et~al.}
\newblock \bibinfo{journal}{\bibinfo{title}{Entangled-photon decision maker}}.
\newblock {\emph{\JournalTitle{Scientific Reports,}}}
  \textbf{\bibinfo{volume}{9(1)}}, \bibinfo{pages}{1--14}
  (\bibinfo{year}{2019}).

\bibitem{Lai}
\bibinfo{author}{Lai, L.}, \bibinfo{author}{El~Gamal, H.}, \bibinfo{author}{H.,
  J.} \& \bibinfo{author}{Poor, H.~V.}
\newblock \bibinfo{journal}{\bibinfo{title}{Cognitive medium access:
  Exploration, exploitation, and competition}}.
\newblock {\emph{\JournalTitle{IEEE Trans. Mob. Comput.}}}
  \textbf{\bibinfo{volume}{10}}, \bibinfo{pages}{239–253}
  (\bibinfo{year}{2010}).

\bibitem{Kim}
\bibinfo{author}{Kim, S.~J.}, \bibinfo{author}{Naruse, M.} \&
  \bibinfo{author}{Aono, M.}
\newblock \bibinfo{journal}{\bibinfo{title}{Harnessing the computational power
  of fluids for optimization of collective decision making}}.
\newblock {\emph{\JournalTitle{Philosophies}}} \textbf{\bibinfo{volume}{1}},
  \bibinfo{pages}{245--260} (\bibinfo{year}{2016}).

\bibitem{steinbrecher2019quantum}
\bibinfo{author}{Steinbrecher, G.~R.}, \bibinfo{author}{Olson, J.~P.},
  \bibinfo{author}{Englund, D.} \& \bibinfo{author}{Carolan, J.}
\newblock \bibinfo{journal}{\bibinfo{title}{Quantum optical neural networks}}.
\newblock {\emph{\JournalTitle{npj Quantum Information}}}
  \textbf{\bibinfo{volume}{5}}, \bibinfo{pages}{60} (\bibinfo{year}{2019}).

\bibitem{saggio2021experimental}
\bibinfo{author}{Saggio, V.} \emph{et~al.}
\newblock \bibinfo{journal}{\bibinfo{title}{Experimental quantum speed-up in
  reinforcement learning agents}}.
\newblock {\emph{\JournalTitle{Nature}}} \textbf{\bibinfo{volume}{591}},
  \bibinfo{pages}{229--233} (\bibinfo{year}{2021}).

\bibitem{flamini2020photonic}
\bibinfo{author}{Flamini, F.} \emph{et~al.}
\newblock \bibinfo{journal}{\bibinfo{title}{Photonic architecture for
  reinforcement learning}}.
\newblock {\emph{\JournalTitle{New Journal of Physics}}}
  \textbf{\bibinfo{volume}{22}}, \bibinfo{pages}{045002}
  (\bibinfo{year}{2020}).

\bibitem{bukov2018reinforcement}
\bibinfo{author}{Bukov, M.} \emph{et~al.}
\newblock \bibinfo{journal}{\bibinfo{title}{Reinforcement learning in different
  phases of quantum control}}.
\newblock {\emph{\JournalTitle{Physical Review X}}}
  \textbf{\bibinfo{volume}{8}}, \bibinfo{pages}{031086} (\bibinfo{year}{2018}).

\bibitem{niu2019universal}
\bibinfo{author}{Niu, M.~Y.}, \bibinfo{author}{Boixo, S.},
  \bibinfo{author}{Smelyanskiy, V.~N.} \& \bibinfo{author}{Neven, H.}
\newblock \bibinfo{journal}{\bibinfo{title}{Universal quantum control through
  deep reinforcement learning}}.
\newblock {\emph{\JournalTitle{npj Quantum Information}}}
  \textbf{\bibinfo{volume}{5}}, \bibinfo{pages}{33} (\bibinfo{year}{2019}).

\bibitem{porotti2019coherent}
\bibinfo{author}{Porotti, R.}, \bibinfo{author}{Tamascelli, D.},
  \bibinfo{author}{Restelli, M.} \& \bibinfo{author}{Prati, E.}
\newblock \bibinfo{journal}{\bibinfo{title}{Coherent transport of quantum
  states by deep reinforcement learning}}.
\newblock {\emph{\JournalTitle{Communications Physics}}}
  \textbf{\bibinfo{volume}{2}}, \bibinfo{pages}{61} (\bibinfo{year}{2019}).

\bibitem{amakasu}
\bibinfo{author}{Amakasu, T.}, \bibinfo{author}{Chauvet, N.},
  \bibinfo{author}{Huant, G.}, \bibinfo{author}{Horisaki, R.} \&
  \bibinfo{author}{Naruse, M.}
\newblock \bibinfo{journal}{\bibinfo{title}{Conflict-free collective stochastic
  decision making by orbital angular momentum of photons through quantum
  interference}}.
\newblock {\emph{\JournalTitle{Scientific Reports,}}}
  \textbf{\bibinfo{volume}{11(1)}}, \bibinfo{pages}{1--13}
  (\bibinfo{year}{2021}).

\bibitem{shinkawa2022conflict}
\bibinfo{author}{Shinkawa, H.} \emph{et~al.}
\newblock \bibinfo{journal}{\bibinfo{title}{Conflict-free joint sampling for
  preference satisfaction through quantum interference}}.
\newblock {\emph{\JournalTitle{Physical Review Applied}}}
  \textbf{\bibinfo{volume}{18}}, \bibinfo{pages}{064018}
  (\bibinfo{year}{2022}).

\bibitem{holzer2000assessing}
\bibinfo{author}{Holzer, H.} \& \bibinfo{author}{Neumark, D.}
\newblock \bibinfo{journal}{\bibinfo{title}{Assessing affirmative action}}.
\newblock {\emph{\JournalTitle{Journal of Economic Literature}}}
  \textbf{\bibinfo{volume}{38}}, \bibinfo{pages}{483--568}
  (\bibinfo{year}{2000}).

\bibitem{bolton2000erc}
\bibinfo{author}{Bolton, G.~E.} \& \bibinfo{author}{Ockenfels, A.}
\newblock \bibinfo{journal}{\bibinfo{title}{Erc: A theory of equity,
  reciprocity, and competition}}.
\newblock {\emph{\JournalTitle{American economic review}}}
  \textbf{\bibinfo{volume}{91}}, \bibinfo{pages}{166--193}
  (\bibinfo{year}{2000}).

\bibitem{blau1994rising}
\bibinfo{author}{Blau, F.~D.} \& \bibinfo{author}{Kahn, L.~M.}
\newblock \bibinfo{journal}{\bibinfo{title}{Rising wage inequality and the us
  gender gap}}.
\newblock {\emph{\JournalTitle{The American Economic Review}}}
  \textbf{\bibinfo{volume}{84}}, \bibinfo{pages}{23--28}
  (\bibinfo{year}{1994}).

\bibitem{shen2013inequality}
\bibinfo{author}{Shen, H.}
\newblock \bibinfo{journal}{\bibinfo{title}{Inequality quantified: Mind the
  gender gap}}.
\newblock {\emph{\JournalTitle{Nature News}}} \textbf{\bibinfo{volume}{495}},
  \bibinfo{pages}{22} (\bibinfo{year}{2013}).

\bibitem{sandel2020tyranny}
\bibinfo{author}{Sandel, M.~J.}
\newblock \emph{\bibinfo{title}{The tyranny of merit: What's become of the
  common good?}} (\bibinfo{publisher}{Penguin UK}, \bibinfo{year}{2020}).

\bibitem{breen2005inequality}
\bibinfo{author}{Breen, R.} \& \bibinfo{author}{Jonsson, J.~O.}
\newblock \bibinfo{journal}{\bibinfo{title}{Inequality of opportunity in
  comparative perspective: Recent research on educational attainment and social
  mobility}}.
\newblock {\emph{\JournalTitle{Annu. Rev. Sociol.}}}
  \textbf{\bibinfo{volume}{31}}, \bibinfo{pages}{223--243}
  (\bibinfo{year}{2005}).

\bibitem{arrieta2020explainable}
\bibinfo{author}{Arrieta, A.~B.} \emph{et~al.}
\newblock \bibinfo{journal}{\bibinfo{title}{Explainable artificial intelligence
  (xai): Concepts, taxonomies, opportunities and challenges toward responsible
  ai}}.
\newblock {\emph{\JournalTitle{Information fusion}}}
  \textbf{\bibinfo{volume}{58}}, \bibinfo{pages}{82--115}
  (\bibinfo{year}{2020}).

\bibitem{Stahl2018}
\bibinfo{author}{Stahl, B.~C.} \& \bibinfo{author}{Wright, D.}
\newblock \bibinfo{journal}{\bibinfo{title}{Ethics and privacy in ai and big
  data: Implementing responsible research and innovation}}.
\newblock {\emph{\JournalTitle{IEEE Security \& Privacy}}}
  \textbf{\bibinfo{volume}{16}}, \bibinfo{pages}{26--33}
  (\bibinfo{year}{2018}).

\bibitem{Yao}
\bibinfo{author}{Yao, A.~M.} \& \bibinfo{author}{Padgett, M.~J.}
\newblock \bibinfo{journal}{\bibinfo{title}{Orbital angular momentum: origins,
  behavior and applications}}.
\newblock {\emph{\JournalTitle{Advances in optics and photonics}}}
  \textbf{\bibinfo{volume}{3}}, \bibinfo{pages}{161--204}
  (\bibinfo{year}{2011}).

\bibitem{Vallone2014}
\bibinfo{author}{Vallone, G.} \emph{et~al.}
\newblock \bibinfo{journal}{\bibinfo{title}{Free-space quantum key distribution
  by rotation-invariant twisted photons}}.
\newblock {\emph{\JournalTitle{Phys. Rev. Lett.}}}
  \textbf{\bibinfo{volume}{113}}, \bibinfo{pages}{060503}
  (\bibinfo{year}{2014}).

\end{thebibliography}



\section*{Author contributions statement}


M.N. and N.C. conceived the project and experimental setup. 
Ho.S. performed mathematical analysis and numerical experiments.
G.B. and J.L. provided support from the experimental perspective.
The manuscript was written by Ho.S., with assistance from M.N., A.R., and Hi.S. It was reviewed by all authors.






\end{document}


\maketitle

\thispagestyle{empty}

\section{Boundary of the feasible zone (Figure 3)}
\subsection{Asymmetric OAM system (Figure 3a)}
The boundary of the feasible pairs of $p_{12}$ and $p_{21}$ in the case of the asymmetric OAM system obtained using Eq. (19) in the main text was derived. 
First, the upper bound of $p_{12}+p_{21}$ was proved to be $(1+(p_{12}-p_{21})^2)/2$.
In particular,
\begin{equation}
    p_{12}+p_{21} \leq \frac{1+(p_{12}-p_{21})^2}{2}.
    \label{OAM_proof0}
\end{equation}
Based on the probabilities listed in Table 1 in the main text,
\begin{align}
    \frac{1+(p_{12}-p_{21})^2}{2} &- (p_{12}+p_{21}) \notag \\ &= \frac{1}{2} (a_2 b_1 - a_1 b_2 -1) (a_2 b_1 - a_1 b_2 +1) \{ 2\alpha\beta (a_2 b_1 + a_1 b_2) -1 \}\{ 2\alpha\beta (a_2 b_1 + a_1 b_2) +1 \}. \label{OAM_proof1} 
\end{align}
Herein, because $a_1$ and $a_2$ satisfy $a_1^2 + a_2^2=1$, $\theta_a$ can be introduced such that $a_1=\cos{\theta_a},~~a_2=\sin{\theta_a}$.
Similarly, $\theta_b$ can be introduced such that $b_1=\cos{\theta_b},~~b_2=\sin{\theta_b}$. 
Meanwhile, with $\theta$, let $\alpha = \sin{\theta}$ and $\beta=\cos{\theta}$.
Therefore,
\begin{equation}
\label{eq:A}
    a_2 b_1 - a_1 b_2 = \sin{\theta_a} \cos{\theta_b} - \cos{\theta_a} \sin{\theta_b} = \sin( \theta_a - \theta_b),
\end{equation}
\begin{equation}
\label{eq:B}
    a_2 b_1 + a_1 b_2 = \sin{\theta_a} \cos{\theta_b} + \cos{\theta_a} \sin{\theta_b} = \sin( \theta_a + \theta_b),
\end{equation}
\begin{equation}
\label{eq:C}
    \alpha \beta = \sin\theta \cos\theta = \frac{1}{2} \sin 2\theta .
\end{equation}
Applying Eqs. \eqref{eq:A} to \eqref{eq:C} into Eq. \eqref{OAM_proof1}, we obtain: 
\begin{equation}
    \frac{1+(p_{12}-p_{21})^2}{2} - (p_{12}+p_{21}) = \frac{1}{2} \{ \sin^2 ( \theta_a - \theta_b) -1 \} \{ \sin^2  2\theta  \sin^2 ( \theta_a + \theta_b) -1 \} .
\end{equation}
Because $0 \leq \sin^2 ( \theta_a - \theta_b) \leq 1$, $ 0 \leq \sin^2  2\theta \leq 1$, $0 \leq \sin^2 ( \theta_a - \theta_b) \leq 1$, 
\begin{equation}
    \sin^2 ( \theta_a - \theta_b) -1 \leq 0, ~~\sin^2  2\theta  \sin^2 ( \theta_a - \theta_b) -1 \leq 0
\end{equation}
holds.
Therefore the following is true, 
\begin{equation}
    0 \leq \frac{1+(p_{12}-p_{21})^2}{2} - (p_{12}+p_{21}) 
\end{equation}
Thus, Eq. \eqref{OAM_proof0} is proven.
As $\frac{1+(p_{12}-p_{21})^2}{2} - (p_{12}+p_{21})$ cannot be negative, $p_{12}+p_{21}$ obviously cannot be larger than $\frac{1+(p_{12}-p_{21})^2}{2}$.
Therefore, the boundary in Figure 3a is proven to be $\frac{1+(p_{12}-p_{21})^2}{2}$.
Equation \eqref{OAM_proof0} holds when $\theta_a,~~\theta_b,~~\theta$ is represented by integers $n,~~m,~~k$, as follows.
\begin{equation}
    \theta_a = \frac{(n+m+1)\pi}{2}, ~~\theta_b=\frac{m-n}{2} \pi,~~\theta=\frac{2k+1}{4}.
\end{equation}
Therefore, the upper bound of $p_{12}+p_{21}$ is proven to be $(1+(p_{12}-p_{21})^2)/2$.
In addition, the lower bound of $p_{12}+p_{21}$ is $0$.
As $p_{12}+p_{21}$ is continuous, $p_{12}+p_{21}$ takes all values from $0$ to $(1+(p_{12}-p_{21})^2)/2$:
\begin{equation}
    0 \leq p_{12}+p_{21} \leq \frac{1+(p_{12}-p_{21})^2}{2}.
    \label{OAM_proof2}
\end{equation}
By considering the condition of $p_{12}$ and $p_{21}$, , evidently: 
\begin{equation}
    p_{12} \geq 0, ~~ p_{21} \geq 0,
    \label{OAM_proof3}
\end{equation}
because $p_{12}$ and $p_{21}$ are probabilities.
Based on the condition of Eq. \eqref{OAM_proof2} and \eqref{OAM_proof3}, the feasible pairs of $p_{12}$ and $p_{21}$ are proven to be the blue-colored areas in Figure 3a in the main text. $\blacksquare$

\subsection{Asymmetric entangled photon decision maker (Figure 3b)}
As proven in the main text that 
\begin{equation}
    p_{12} = \frac{\sin^2\beta_Y \cos^2\alpha_X}{2},  ~~p_{21}=\frac{\sin^2\beta_X \cos^2\alpha_Y}{2}.
\end{equation}
Because $\alpha_X$, $\alpha_Y$, $\beta_X$, and $\beta_Y$ are independent, 
\begin{equation}
    0 \leq p_{12} \leq \frac{1}{2}, ~~ 0 \leq p_{21} \leq \frac{1}{2}.
    \label{etg_3b}
\end{equation}
Provided that Eq. \eqref{etg_3b} holds, $p_{12}$ and $p_{21}$ can take any value owing to the continuity of trigonometric functions.
Therefore, the boundary by the asymmetric entangled photon decision maker in Figure 3b is proven. $\blacksquare$

\subsection{OAM attenuation (Figure 3c)}
As explained in the main text that
\begin{equation}
    p_{12} = \left( a_1^2 b_2^2 + a_2^2 b_1^2 - 2 a_1 a_2 b_1 b_2 \cos(\theta_1-\theta_2) \right) d_{aX}^2 d_{bY}^2,  
    ~~p_{21}=\left( a_2^2 b_1^2 + a_1^2 b_2^2 - 2 a_1 a_2 b_1 b_2 \cos(\theta_1-\theta_2) \right) d_{aY}^2 d_{bX}^2.
\end{equation}
They have $a_1^2 b_2^2 + a_2^2 b_1^2 - 2 a_1 a_2 b_1 b_2 \cos(\theta_1-\theta_2)$ in common; however,  the attenuation rates multiplied are different.
As the following is true:
\begin{equation}
    0 \leq a_1^2 b_2^2 + a_2^2 b_1^2 - 2 a_1 a_2 b_1 b_2 \cos(\theta_1-\theta_2) \leq \frac{1}{4},
\end{equation}
and owing to the fact that $d_{aX}$, $d_{aY}$, $d_{bX}$, and $d_{bY}$ are independent, the boundary by the asymmetric entangled photon decision maker in Figure 3c is proven. $\blacksquare$

\section{Boundary of asymmetric ratio and loss plus conflict probability (Figure 4)}
\subsection{Asymmetric OAM system (Figure 4a)}
In case of Figure 3a (in the main text), points on the boundary have the maximum $p_{12}+p_{21}$, provided $p_{21}/p_{12}$ is consistent.
The values of loss probability plus conflict probability are minimum on the boundary for the same $p_{21}/p_{12}$.
Therefore, if the boundary in Figure 3a can be converted to that in Figure 4a (in the main text), then the boundary in Figure 4a is mathematically derived.
This conversion is shown as follows.

Let $y$ be the ratio of $p_{12}$ and $p_{21}$: $p_{21} = y p_{12}$.
Let $x$ be loss probability plus conflict probability on the boundary in Figure 3b.
Here, $x$ is $p_{12}+p_{21}$ on the intersection of the boundary in Figure 3a and line of $p_{21} = y p_{12}$.
We refer to this intersection as $V_1$.
On $V_1$,
\begin{equation}
    2(1+y) p_{12} = 1+(1-y)^2 p_{12}^2,
\end{equation}
\begin{equation}
    (1-y)^2 p_{12}^2 -2(1+y)p_{12} +1=0,
\end{equation}
\begin{equation}
    p_{12} = \frac{1+y \pm 2\sqrt{y}}{(1-y)^2} =\frac{(1\pm\sqrt{y})^2}{(1-y)^2}. \label{OAM_pm}
\end{equation}
If 
\begin{equation}
    p_{12} = \frac{(1 + \sqrt{y})^2}{(1-y)^2},
\end{equation}
then, 
\begin{equation}
    x = 1-(p_{12}+p_{21}) = 1 - \frac{y+1}{(\sqrt{y}-1)^2} = \frac{- 2\sqrt{y}}{(\sqrt{y}-1)^2} < 0.
\end{equation}
This is strange because $x$ should satisfy $0 \leq x \leq 1$.
Therefore, the sign in Eq. \eqref{OAM_pm} is determined:
\begin{equation}
    p_{12} = \frac{(1 - \sqrt{y})^2}{(1-y)^2}.
\end{equation} 
Hence, 
\begin{equation}
    x = 1- (p_{12}+p_{21}) = 1- \frac{y+1}{(\sqrt{y}+1)^2} =\frac{2\sqrt{y}}{(\sqrt{y}+1)^2}. \label{OAM_x_of_y}
\end{equation}
By solving Eq. \eqref{OAM_x_of_y}, we obtain:
\begin{equation}
    y=\begin{cases}
        \cfrac{(1+\sqrt{1-2x})^2}{(1-\sqrt{1-2x})^2} ~~& \text{when } y \ge 1 ,\\
        \\  
        \cfrac{(1-\sqrt{1-2x})^2}{(1+\sqrt{1-2x})^2}~~& \text{when } y \le 1. 
    \end{cases}
\end{equation} 
Thus, Eq. (20) in the main text is derived. $\blacksquare$

\subsection{Asymmetric entangled photon decision maker (Figure 4b)}
In Figure 3b, points on the boundary have the maximum $p_{12}+p_{21}$ and minimum loss probability plus conflict probability for the same $p_{21}/p_{12}$, as in the previous subsection.
If the boundary in Figure 3b can be converted to the boundary in Figure 4b, then that in Figure 4a is mathematically derived.
This conversion is shown as follows.

Let $y$ be the ratio of $p_{12}$ and $p_{21}$: $p_{21} = y p_{12}$.
Let $x$ be loss probability plus conflict probability on the boundary in Figure 3b.
First, consider the case where $y$ is equal to or greater than $1$.
In this case,
\begin{equation}
    p_{12} = \frac{1}{2y},~~p_{21}=\frac{1}{2}.
\end{equation}
Therefore,
\begin{align}
    x &= 1-(p_{12}+p_{21}) = 1-\left( \frac{1}{2y} + \frac{1}{2} \right) \\
    &= \frac{y-1}{2y}. \label{etg_4b_1}
\end{align}
By solving Eq. \eqref{etg_4b_1} about $y$, we obtain:
\begin{equation}
    y = \frac{1}{1-2x}.
\end{equation}
Next, when  $y$ is equal to or smaller than $1$,
\begin{equation}
    p_{12} = \frac{1}{2},~~p_{21}=\frac{y}{2}.
\end{equation}
Therefore, 
\begin{align}
    x &= 1-(p_{12}+p_{21}) = 1-\left( \frac{1}{2} + \frac{y}{2} \right) \\
    &= \frac{1-y}{2}. \label{etg_4b_2}
\end{align}
By solving Eq. \eqref{etg_4b_2} about $y$, we obtain:
\begin{equation}
    y = 1-2x.
\end{equation}
Therefore, the mathematical formula of the boundary is expressed as:
\begin{equation}
    y=\begin{cases}
    \cfrac{1}{1-2x} ~~& \text{when } y \ge 1, \\
    \\  
    1-2x~~& \text{when } y \le 1. ~~
    \end{cases}
\end{equation}
This corresponds to Eq. (35) in the main text. $\blacksquare$

\subsection{OAM attenuation (Figure 4c)}
Next, the conversion from the boundary in Figure 3c to that in Figure 4c is as follows.
Let $y$ be the ratio of $p_{12}$ and $p_{21}$: $p_{21} = y p_{12}$.
Let $x$ be loss probability plus conflict probability on the boundary in Figure 3b.
First, consider the case where $y$ is equal to or greater than $1$.
In this case,
\begin{equation}
    p_{12} = \frac{1}{4y},~~p_{21}=\frac{1}{4}.
\end{equation}
Therefore,
\begin{align}
    x &= 1-(p_{12}+p_{21}) = 1-\left( \frac{1}{4y} + \frac{1}{4} \right) \\
    &= \frac{3y-1}{4y}. \label{att_4c_1}
\end{align}
By solving Eq. \eqref{att_4c_1} about $y$, we obtain:
\begin{equation}
    y = \frac{1}{3 - 4x}.
\end{equation}
Next, when  $y$ is equal to or smaller than $1$,
\begin{equation}
    p_{12} = \frac{1}{4},~~p_{21}=\frac{y}{4}.
\end{equation}
Therefore, 
\begin{align}
    x &= 1-(p_{12}+p_{21}) = 1-\left( \frac{1}{4} + \frac{y}{4} \right) \\
    &= \frac{3-y}{4}. \label{att_4c_2}
\end{align}
By solving Eq. \eqref{att_4c_2} about $y$, we obtain:
\begin{equation}
    y = 3 - 4x.
\end{equation}
Therefore, the mathematical formula of the boundary is expressed as:
\begin{equation}
    y=\begin{cases}
    \cfrac{1}{3 - 4x} ~~& \text{when } y \ge 1, \\
    \\  
    3 - 4x ~~& \text{when } y \le 1, 
    \end{cases}
\end{equation}
Thus, Eq. (39) in the main text is derived. $\blacksquare$